\documentclass[zpreprint,zbstnp]{zeus_paper}
\prepnum{{DESY--12--144}}
\date{August 2012} 
\title{Production of the excited charm mesons $\boldsymbol{D_1}$ and $\boldsymbol{D^*_2}$ at HERA }
\author{ZEUS Collaboration}
\newcommand{\EDMtitle}{
\abstract{
The production of the excited charm mesons $D_1(2420)$ and                           
                         $D^*_2(2460)$ in $e p$ collisions
has been measured with the ZEUS detector at HERA using an integrated luminosity of $373\pbi$.
The masses of the neutral and charged states, the widths  of the neutral 
states, and the helicity parameter of $D_1(2420)^{0}$   were determined and compared with  
other measurements
and
with theoretical expectations.         
The measured helicity parameter of the  $D^0_1$  allows for some
mixing of $S$- and $D$-waves in its decay to $D^{*\pm}\pi^{\mp}$. The result is also consistent
with  a pure $D$-wave decay.
Ratios of branching fractions of the two decay modes of the $D^*_2(2460)^0$ and $D^*_2(2460)^{\pm}$ 
                                                                 states were measured and compared with
previous measurements.    
The fractions of charm quarks hadronising into $D_1$ and $D^{*}_2$ were measured and are consistent         
                                                                                        with those obtained in
$e^+ e^-$ annihilations.
}
\makezeustitle \newpage}

\newcommand{\bran}{\mbox{$\cal B$}}
\newcommand{\fr}{\mbox{$\cal F$}}

\newcommand{\dsp}        {\mbox{$D^{\ast +}$}}

\newcommand{\dz}         {\mbox{$D^{0}$}}

\newcommand{\dc}         {\mbox{$D^+$}}

\newcommand{\done}       {\mbox{$D_1^{0}$}}
\newcommand{\dtwo}       {\mbox{$D_2^{\ast 0}$}}

\newcommand{\fcdone}     {\mbox{$f(c \rightarrow D_1^0)$}}
\newcommand{\fcdtwo}     {\mbox{$f(c \rightarrow D_2^{\ast 0})$}}

\newcommand{\fcdz}       {\mbox{$f(c \rightarrow D^0)$}}
\newcommand{\fcdc}       {\mbox{$f(c \rightarrow D^+)$}}

\newcommand{\fcds}       {\mbox{$f(c \rightarrow D^{\ast +})$}}

\def\dsk3pi{ {\dsp}~\rightarrow~\dz~\pi^{+}_{s}\rightarrow~(K^{-}~\pi^{+}~\pi^{+}~\pi^{-})~\pi^{+}_{s} }

\def\et10t{ E_T^{\theta > 10^\circ}}
\def\etw10{ E_T^{\theta > 10}}

\newcommand{\eVdist}{\kern-0.06667em}

\newcommand{\pbi}{\,\text{pb}^{-1}}

\newcommand{\degree}{\ensuremath{^{\circ}}}
\newcommand{\tesla}{\ensuremath{\,\text{T}}}
\newcommand{\micron}{\ensuremath{\,\upmu\text{m}}}
\newcommand{\GeV}{{\text{Ge}\eVdist\text{V\/}}}

\newcommand{\DStarTwoProngMass} {{$145.400\pm0.003$~MeV}\xspace}
\newcommand{\DStarFourProngMass} {{$145.420\pm0.003$~MeV}\xspace}
\newcommand{\DZeroMass} {{$1864.1\pm0.1$~MeV}\xspace}
\newcommand{\DChargeMass} {{$1869.1\pm 0.1$~MeV}\xspace}

\newcommand{\DStarTwoProngYield}    {{$64988 \pm 430$}\xspace}
\newcommand{\DStarFourProngYield}   {{$24441 \pm 310$}\xspace}
\newcommand{\DZeroYield}            {{$145740 \pm 2944$}\xspace}
\newcommand{\DChargeYield}          {{$39283 \pm 452$}\xspace}

\newcommand{\DChargeOneYield}                {{$759 \pm  183$}\xspace}
\newcommand{\DChargeOneMass}                 {{$2421.9\pm 4.7^{+3.4}_{-1.2}$}\xspace}
\newcommand{\DChargeStarTwoYieldDStarZeroPi} {{$634 \pm 223$}\xspace}
\newcommand{\DChargeStarTwoYieldDZeroPi}     {{$737 \pm 164$}\xspace}
\newcommand{\DChargeStarTwoMass}             {{$2460.6 \pm 4.4^{+3.6}_{-0.8}$}\xspace}

\newcommand{\DZeroOneYield}            {{$2732 \pm 285$}\xspace}
\newcommand{\DZeroOneMass}             {{$2423.1 \pm 1.5{}^{+ 0.4}_{- 1.0}$}\xspace}

\newcommand{\DZeroOneWidth}            {{$38.8\pm 5.0{}^{+ 1.9}_{- 5.4}$}\xspace}
\newcommand{\DZeroOneWidthStatSyst}    {{$38.8 \pm 5.0\rm{(stat.)}{}^{+ 1.9}_{- 5.4}\rm{(syst.)}$}\xspace}
\newcommand{\DZeroOneHelicity}         {{$7.8{}^{+6.7}_{-2.7}{}^{+4.6}_{-1.8}$}\xspace}
\newcommand{\DZeroOneHelicityStatSyst} {{$7.8{}^{+6.7}_{-2.7}\rm{(stat.)}{}^{+4.6}_{-1.8}\rm{(syst.)}$}\xspace}

\newcommand{\DZeroStarTwoYieldDStarPi}  {{$1798 \pm 293$}\xspace}
\newcommand{\DZeroStarTwoYieldDPlusPi}  {{$521 \pm  88$}\xspace}
\newcommand{\DZeroStarTwoMass}          {{$2462.5 \pm 2.4{}^{+ 1.3}_{- 1.1}$}\xspace}

\newcommand{\DZeroStarTwoWidth}         {{$46.6 \pm 8.1{}^{+ 5.9}_{- 3.8}$}\xspace}
\newcommand{\DZeroStarTwoWidthStatSyst} {{$46.6 \pm 8.1\rm{(stat.)}{}^{+ 5.9}_{- 3.8}\rm{(syst.)}$}\xspace}
\newcommand{\DZeroStarTwoHelicity}      {{$-1.16 \pm 0.35$}\xspace}

\newcommand{\DTwoFourZeroZeroRatio}     {{$1.1 \pm 1.1$}\xspace}
\newcommand{\DTwoFourThreeZeroRatio}    {{$1.0$ fixed }\xspace}
\newcommand{\FeeddownRatio}             {{$0.3\pm 0.4$ }\xspace}

\newcommand{\DZeroOneFragmentationStatSyst}              {\mbox{$2.9\pm0.5\rm{(stat.)}\pm0.5\rm{(syst.)}\,\%$}\xspace}

\newcommand{\DZeroOneExtrapolationDStarStatSyst}         {\mbox{$8.5\pm1.4\rm{(stat.)}{}^{+1.2}_{-1.6}\rm{(syst.)}\,\%$}\xspace}

\newcommand{\DZeroStarTwoFragmentationStatSyst}          {\mbox{$3.9\pm0.9\rm{(stat.)}{}^{+0.8}_{-0.6}\rm{(syst.)}\,\%$}\xspace}

\newcommand{\DZeroStarTwoExtrapolationDStarStatSyst}     {\mbox{$4.7\pm1.3\rm{(stat.)}{}^{+1.2}_{-0.8}\rm{(syst.)}\,\%$}\xspace}

\newcommand{\DZeroStarTwoExtrapolationDPlusStatSyst}     {\mbox{$6.7\pm2.4\rm{(stat.)}{}^{+1.5}_{-1.1}\rm{(syst.)}\,\%$}\xspace}

\newcommand{\DZeroStarTwoBranchingStatSyst}              {\mbox{$1.4 \pm 0.3\rm{(stat.)}{\pm 0.3}\rm{(syst.)}$}\xspace}

\newcommand{\DExcitedNeutralFragmentationSumStatSyst}    {\mbox{$6.8\pm1.0\rm{(stat.)}{}^{+0.9}_{-0.8}\rm{(syst.)}\,\%$}\xspace}

\newcommand{\DExcitedNeutralFragmentationRatioStatSyst}  {\mbox{$0.8\pm0.2\rm{(stat.)}\pm0.2\rm{(syst.)}$}\xspace}

\newcommand{\fcdoneplus}     {\mbox{$f(c \rightarrow D_1^+)$}}
\newcommand{\fcdtwoplus}     {\mbox{$f(c \rightarrow D_2^{\ast +})$}}

\newcommand{\DPlusStarTwoBranchingStatSyst}              {\mbox{$1.1 \pm 0.4\rm{(stat.)}{}^{+0.3}_{-0.2}\rm{(syst.)}$}\xspace}

\newcommand{\DPlusStarTwoBranchingBABARStyleStatSyst}              {\mbox{$0.52^{+0.08}_{-0.13}\rm{(stat.)}\pm0.05\rm{(syst.)}$}\xspace}

\newcommand{\DPlusOneFragmentationStatSyst}              {\mbox{$4.6\pm1.8\rm{(stat.)}{}^{+2.0}_{-0.3}\rm{(syst.)}\,\%$}\xspace}

\newcommand{\DPlusStarTwoFragmentationStatSyst}          {\mbox{$3.2\pm0.8\rm{(stat.)}{}^{+0.5}_{-0.2}\rm{(syst.)}\,\%$}\xspace}

\newcommand{\DPlusOneExtrapolationDZeroStarStatSyst}     {\mbox{$5.4\pm2.1\rm{(stat.)}{}^{+2.3}_{-0.3}\rm{(syst.)}\,\%$}\xspace}

\newcommand{\DPlusStarTwoExtrapolationDZeroStarStatSyst} {\mbox{$1.8\pm0.9\rm{(stat.)}{}^{+0.5}_{-0.3}\rm{(syst.)}\,\%$}\xspace}

\newcommand{\DPlusStarTwoExtrapolationDZeroStatSyst}     {\mbox{$2.0\pm0.5\rm{(stat.)}{}^{+0.4}_{-0.2}\rm{(syst.)}\,\%$}\xspace}

\newcommand{\DExcitedChargedFragmentationSumStatSyst}    {\mbox{$7.8\pm2.0\rm{(stat.)}{}^{+2.0}_{-0.4}\rm{(syst.)}\,\%$}\xspace}

\newcommand{\DExcitedChargedFragmentationRatioStatSyst}  {\mbox{$1.4\pm0.7\rm{(stat.)}{}^{+0.7}_{-0.1}\rm{(syst.)}$}\xspace}

\newcommand{\tableone} {{
\large{
\begin{table}[h]
\begin{center}
\begin{tabular}{|c|c|c|}
\hline Variable &            
 $D^0\to K^-\pi^+$ &  
 $D^0\to K^-\pi^+\pi^+\pi^-$    
                 \\
 \hline\hline
 $p_T(K)$~(GeV)  & $ > 0.45$ &  $ > 0.3$  
    \\
 \hline
 $p_T(\pi)$~(GeV)  & $ > 0.45$ &  $ > 0.3$  
    \\
 \hline
 $p_T(\pi_s)$~(GeV)  & $ > 0.1 $ &  $ > 0.1$  
    \\
 \hline
 $p_T(D^{*+})$~(GeV)  & $ > 1.5 $ &  $ > 3$  
    \\
 \hline
 $|\eta(D^{*+})|$ & $ < 1.6 $ &  $ < 1.6 $
    \\
 \hline
 $p_T(D^{*+})/E_{\perp}^{\theta > 10^{\circ}}$ & $ > 0.12 $ &  $ > 0.18$  
    \\
\hline
$M(D^0)$~(GeV) for  & $1.83 - 1.90$ & $1.84 - 1.89$                                   
   \\
$p_T(D^{*+}) < 3.25$~GeV &  &
   \\
\hline
$M(D^0)$~(GeV) for  & $1.82 - 1.91$ & $1.84 - 1.89$                                     
   \\
$3.25 < p_T(D^{*+}) < 5$~GeV &  &
   \\
\hline
$M(D^0)$~(GeV) for  & $1.81 - 1.92$ & $1.84 - 1.89$                                   
   \\

$5 < p_T(D^{*+}) < 8$~GeV  &  &
   \\
\hline
$M(D^0)$~(GeV) for  & $1.80 - 1.93$ & $1.84 - 1.89$                                   
   \\
$p_T(D^{*+}) > 8$~GeV &  &
   \\
 
\hline
 
\end{tabular}
 
\end{center}
\caption{Cuts on $D^{*+}\to D^0\pi^+_s$ candidates for the decay channels $D^0\to K^-\pi^+$ and  $D^0\to K^-\pi^+\pi^+\pi^-$.}
\label{tab1}
\end{table}
}
 }}

\newcommand{\tabletwo} {{
{
\begin{table}[h]
\begin{center}
\begin{tabular}{|c|c|c|c|}\hline
                       & HERA II        &       HERA I &        PDG         \\
 \hline\hline
 $N(D^0_1\to D^{*+}\pi)$    & \DZeroOneYield  & $3110\pm 340$ &   \\                               \hline
 $N(D^{*0}_2\to D^{*+}\pi)$ & \DZeroStarTwoYieldDStarPi & $ 870\pm 170$  &  
    \\
 \hline
 $N(D^{*0}_2\to D^+\pi)$    &\DZeroStarTwoYieldDPlusPi  $(S(D^+) > 3)$ & $ 690\pm 160  $ &  
    \\
\hline
$M(D^0_1)$,~MeV & \DZeroOneMass & $2420.5\pm 2.1\pm 0.9$ & $2421.3\pm 0.6$                           
   \\
\hline
$\Gamma(D^0_1)$,~MeV  & \DZeroOneWidth & $53.2\pm 7.2^{+3.3}_{-4.9}$& $27.1\pm 2.7$                         
   \\
\hline
$h(D^0_1)$  & \DZeroOneHelicity & $5.9^{+3.0}_{-1.7}{}^{+2.4}_{-1.0}$&                    
   \\
\hline
$M(D^{*0}_2)$,~MeV  & \DZeroStarTwoMass & $2469.1\pm 3.7^{+1.2}_{-1.3}$ & $2462.6\pm 0.7$    
   \\
\hline
$\Gamma(D^{*0}_2)$,~MeV  & \DZeroStarTwoWidth& $43$ fixed   & $49.0\pm 1.4$                       
   \\
\hline
$h(D^{*0}_2)$  & $-1$ fixed &    $-1$ fixed     &
   \\
\hline
$D_1(2430)^0/D^0_1 $   &   \DTwoFourThreeZeroRatio  &  $1.0$ fixed   &
   \\
\hline
$D^*_0(2400)^0/D^{*0}_2 $  &  \DTwoFourZeroZeroRatio  & $1.7$ fixed    &
   \\
\hline
Feed-downs/$D^{*0}_2 $    &    \FeeddownRatio &  &
   \\
\hline
\end{tabular}
\end{center}
\caption{Results of the simultaneous fit for the yields (N), masses (M),  widths ($\Gamma$) 
 and helicity parameters (h) of the $D^0_1$ and $D^{*0}_2$ mesons, for the ratios of the wide states 
 $D_1(2430)^0$ and $D^*_0(2400)^0$ to the narrow states $D^0_1$ and $D^{*0}_2$, and for the ratio of
the feed-down (see text) to the $D^{*0}_2\to D^+\pi^-$. The first uncertainties are statistical and the second
  are systematic. The results (HERA II) are compared to earlier ZEUS results at                                              
                     HERA I~\protect\cite{dsshera1} and to the PDG~\protect\cite{PDG11}.}
\label{tab2}
\end{table}
}
}}
 
\newcommand{\tablethree} {{
\large{
\begin{table}[h]
\begin{center}

\begin{tabular}{|c|c|c|}

\hline     &                         
             HERA II &  
       PDG         \\
 \hline\hline
 $N(D^+_1\to D^{*0}\pi^+)$  &\DChargeOneYield &                    
    \\                              
 \hline
 $N(D^{*+}_2\to D^{*0}\pi^+)$  & \DChargeStarTwoYieldDStarZeroPi &  
    \\
 \hline
 $N(D^{*+}_2\to D^0\pi^+)$  &  \DChargeStarTwoYieldDZeroPi &  
    \\
\hline
$M(D^+_1)$,~MeV &  \DChargeOneMass & $2423.4\pm 3.1$                           
   \\
\hline
$\Gamma(D^+_1)$,~MeV &  $25$ fixed & $25\pm 6$                         
   \\
\hline
$h(D^+_1)$ &  $3.0$ fixed &                    
   \\
\hline
$M(D^{*+}_2)$,~MeV & \DChargeStarTwoMass & $2464.4\pm 1.9$    
   \\
\hline
$\Gamma(D^{*+}_2)$,~MeV  &  $37$ fixed                  & $37\pm 6$                       
   \\
\hline
$h(D^{*+}_2)$ &  $-1.0$ fixed &                    
   \\
\hline 
\end{tabular}
 
\end{center}
\caption{
Results of the fit for the yields (N), masses (M), 
widths ($\Gamma$) and helicity parameters (h) of the $D^+_1$ and
$D^{*+}_2$ mesons.                                             
                The first uncertainties are statistical and the second
 are systematic. The results are compared to those of the PDG~\protect\cite{PDG11}.
                       }
\label{tab3}
\end{table}
}
 }}

\newcommand{\tablefour} {{
\large{
\begin{table}[h]
\begin{center}
\begin{tabular}{|c|c|c|c|c|c|c|c|} \hline
& total,\% &$\delta_1,\%$&$\delta_2,\%$&$\delta_3,\%$&$\delta_4,\%$&$\delta_6,\%$&$\delta_7,\%$\\\hline
\hline
$\fr^{\rm extr}_{D_{1_{}}^{0}\rightarrow D^{*+}\pi^-/D^{*+}}$&
$^{+19.2}_{-14.5}$&$^{+16.4}_{-12.2}$&$^{+ 6.7}_{- 0.0}$&$^{+ 3.4}_{- 7.5}$&$^{+ 0.3}_{- 0.0}$&$^{+ 1.5}_{- 2.0}$&$^{+ 6.5}_{- 0.0}$
\\\hline
$\fr^{\rm extr}_{D_{2_{}}^{*0}\rightarrow D^{*+}\pi^-/D^{*+}}$&
$^{+13.5}_{-18.2}$&$^{+11.9}_{-12.9}$&$^{+ 3.7}_{- 5.0}$&$^{+ 1.2}_{-11.8}$&$^{+ 4.9}_{- 0.0}$&$^{+ 0.9}_{- 1.5}$&$^{+ 0.1}_{- 0.0}$
\\\hline
$\fr^{\rm extr}_{D_{2_{}}^{*0}\rightarrow D^{+}\pi^-/D^{+}}$&
$^{+25.2}_{-17.3}$&$^{+18.6}_{- 7.8}$&$^{+11.9}_{- 0.0}$&$^{+ 5.4}_{-15.4}$&$^{+ 1.0}_{- 0.0}$&$^{+ 0.5}_{- 0.8}$&$^{+10.7}_{- 0.0}$ 
\\\hline
$\frac{\bran_{D_{2_{}}^{*0}\rightarrow D^{+} \pi^-}}{\bran_{D_{2_{}}^{*0} \rightarrow D^{*+} \pi^-}}$&
$^{+20.1}_{-19.5}$&$^{+ 9.9}_{-13.5}$&$^{+ 0.0}_{- 4.7}$&$^{+ 9.6}_{- 3.3}$&$^{+ 0.0}_{- 0.7}$&$^{+ 2.3}_{- 2.5}$&$^{+14.4}_{-12.7}$
\\\hline
$f(c \rightarrow D_1^{0})$&
$^{+15.8}_{-18.6}$&$^{+11.9}_{-12.9}$&$^{+ 3.7}_{- 5.0}$&$^{+ 1.2}_{-11.8}$&$^{+ 4.9}_{- 0.0}$&$^{+ 0.9}_{- 1.5}$&$^{+ 8.1}_{- 3.6}$
\\\hline
$f(c \rightarrow D_2^{\ast 0})$&
$^{+22.4}_{-15.1}$&$^{+16.1}_{- 9.1}$&$^{+ 8.9}_{- 0.0}$&$^{+ 4.0}_{-10.7}$&$^{+ 0.6}_{- 0.0}$&$^{+ 0.6}_{- 1.0}$&$^{+12.2}_{- 5.3}$
\\\hline
\end{tabular}
\caption{
    Total and $\delta_1$-$\delta_7$ (see text)
systematic uncertainties
for extrapolated fractions, for ratios of the dominant branching fractions and
for fragmentation
fractions of the $\done$ and $\dtwo$ mesons.
}
\label{tab:syst_rf}
\end{center}
\end{table}
} 
}}

\newcommand{\tablefive} {{
\large{
\begin{table}[h]
\begin{center}
\begin{tabular}{|c|c|c|c|c|c|c|c|c|} \hline
& total,\% &$\delta_1,\%$&$\delta_2,\%$&$\delta_3,\%$&$\delta_4,\%$&$\delta_5,\%$&$\delta_6,\%$&$\delta_7,\%$\\\hline
\hline
$\fr^{\rm extr}_{D_{1_{}}^{+}\rightarrow D^{*0}\pi^+/D^{0}}$&
$^{+42.6}_{- 6.1}$&$^{+30.5}_{- 0.0}$&$^{+18.3}_{- 0.0}$&$^{+ 3.7}_{- 2.6}$&$^{+ 0.0}_{- 0.0}$&$^{+22.2}_{- 0.0}$&$^{+ 1.8}_{- 5.2}$&$^{+ 6.0}_{- 1.9}$
\\\hline
$\fr^{\rm extr}_{D_{2_{}}^{*+}\rightarrow D^{*0}\pi^+/D^{0}}$&
$^{+24.6}_{-14.8}$&$^{+14.7}_{- 1.3}$&$^{+ 6.3}_{- 2.4}$&$^{+ 1.2}_{- 7.9}$&$^{+ 0.0}_{- 0.0}$&$^{+13.5}_{- 4.6}$&$^{+ 3.5}_{- 4.0}$&$^{+12.5}_{-10.5}$
\\\hline
$\fr^{\rm extr}_{D_{2_{}}^{*+}\rightarrow D^{0}\pi^+/D^{0}}$&
$^{+18.0}_{- 8.0}$&$^{+13.4}_{- 0.8}$&$^{+ 5.6}_{- 4.3}$&$^{+ 0.2}_{- 5.2}$&$^{+ 0.0}_{- 0.0}$&$^{+ 3.6}_{- 0.0}$&$^{+ 1.6}_{- 1.4}$&$^{+ 9.8}_{- 3.9}$
\\\hline
$\frac{\bran_{D_{2_{}}^{*+} \rightarrow D^{0} \pi^+}}{\bran_{D_{2_{}}^{*+} \rightarrow D^{*0} \pi^+}}$&
$^{+23.8}_{-19.1}$&$^{+10.5}_{- 8.5}$&$^{+ 8.3}_{-10.0}$&$^{+ 7.0}_{- 4.7}$&$^{+ 0.0}_{- 0.0}$&$^{+ 6.9}_{- 9.1}$&$^{+ 2.7}_{- 1.9}$&$^{+16.9}_{- 9.3}$
\\\hline
$f(c \rightarrow D_1^{+})$&
$^{+42.7}_{- 7.3}$&$^{+30.5}_{- 0.0}$&$^{+18.3}_{- 0.0}$&$^{+ 3.7}_{- 2.6}$&$^{+ 0.0}_{- 0.0}$&$^{+22.2}_{- 0.0}$&$^{+ 1.8}_{- 5.2}$&$^{+ 7.1}_{- 4.4}$
\\\hline
$f(c \rightarrow D_2^{\ast +})$&
$^{+16.7}_{- 7.1}$&$^{+12.0}_{- 0.0}$&$^{+ 1.8}_{- 0.0}$&$^{+ 0.5}_{- 5.4}$&$^{+ 0.0}_{- 0.0}$&$^{+ 8.2}_{- 1.2}$&$^{+ 2.5}_{- 2.7}$&$^{+ 7.7}_{- 3.6}$
\\\hline
\end{tabular}
\caption{
    Total and $\delta_1$-$\delta_7$ (see text)
systematic uncertainties
for extrapolated fractions, for ratios of the dominant branching fractions and
for fragmentation
fractions of the $D^+_1$ and $D^{*+}_2$ mesons.
}
\label{tab:syst_charged}
\end{center}
\end{table}
}
 }}

\newcommand{\tablesix} {{
\large{
\begin{table}[h]
\begin{center}
\begin{tabular}{|c|c|c|c|c|c|c|} \hline
& total &$\delta_1$&$\delta_2$&$\delta_3$&$\delta_4$\\\hline
\hline
$M(\done)$,~MeV     &
$^{+ 0.4}_{- 1.0}$&$^{+ 0.4}_{- 0.3}$&$^{+ 0.0}_{- 0.8}$&$^{+ 0.1}_{- 0.5}$&$^{+ 0.1}_{- 0.1}$
\\\hline
$M(\dtwo)$,~MeV     &
$^{+ 1.3}_{- 1.1}$&$^{+ 0.9}_{- 0.9}$&$^{+ 0.9}_{- 0.5}$&$^{+ 0.2}_{- 0.2}$&$^{+ 0.0}_{- 0.1}$
\\\hline
$\Gamma(\done)$,~MeV&
$^{+ 1.9}_{- 5.4}$&$^{+ 1.6}_{- 2.3}$&$^{+ 0.0}_{- 1.6}$&$^{+ 1.0}_{- 4.5}$&$^{+ 0.0}_{- 0.0}$
\\\hline
$\Gamma(\dtwo)$,~MeV&
$^{+ 5.9}_{- 3.8}$&$^{+ 4.0}_{- 3.5}$&$^{+ 0.1}_{- 0.2}$&$^{+ 4.3}_{- 1.7}$&$^{+ 0.0}_{- 0.0}$
\\\hline
$h(\done)$           &
$^{+ 4.6}_{- 1.8}$&$^{+ 3.1}_{- 1.3}$&$^{+ 2.4}_{- 0.3}$&$^{+ 2.3}_{- 1.3}$&$^{+ 0.1}_{- 0.1}$
\\\hline
\end{tabular}
\caption{
    Total and $\delta_1$-$\delta_4$
(see text)
systematic uncertainties
for the
mass, width and helicity
parameters of the neutral excited charm mesons.
}
\label{tab:syst_mg}
\end{center}
\end{table}
}
 
 }}

\newcommand{\tableseven} {{
\large{
\begin{table}[hbt]
\begin{center}
\begin{tabular}{|c|c|c|c|c|c|c|} \hline
& total&$\delta_1$&$\delta_2$&$\delta_3$&$\delta_4$&$\delta_5$\\\hline
\hline
$M(D^+_1)$,~MeV       &
$^{+  3.4}_{-  1.2}$&$^{+  3.2}_{-  0.1}$&$^{+  0.0}_{-  0.7}$&$^{+  0.6}_{-  0.1}$&$^{+  0.1}_{-  0.1}$&$^{+  0.6}_{-  0.9}$
\\\hline
$M(D^{*+}_2)$,~MeV    &
$^{+  3.7}_{-  0.8}$&$^{+  1.7}_{-  0.5}$&$^{+  3.1}_{-  0.0}$&$^{+  0.4}_{-  0.2}$&$^{+  0.1}_{-  0.1}$&$^{+  0.9}_{-  0.6}$
\\\hline 
\end{tabular}
\caption{
    Total and $\delta_1$-$\delta_5$
(see text)
systematic uncertainties
for the
mass, width and helicity
parameters of the charged excited charm mesons.
}
\label{tab:syst_cha}
\end{center}
\end{table}
}
}}

\newcommand{\figureone}{
\begin{figure}[ht]
\vfill
\begin{center}
\includegraphics[scale=0.40]{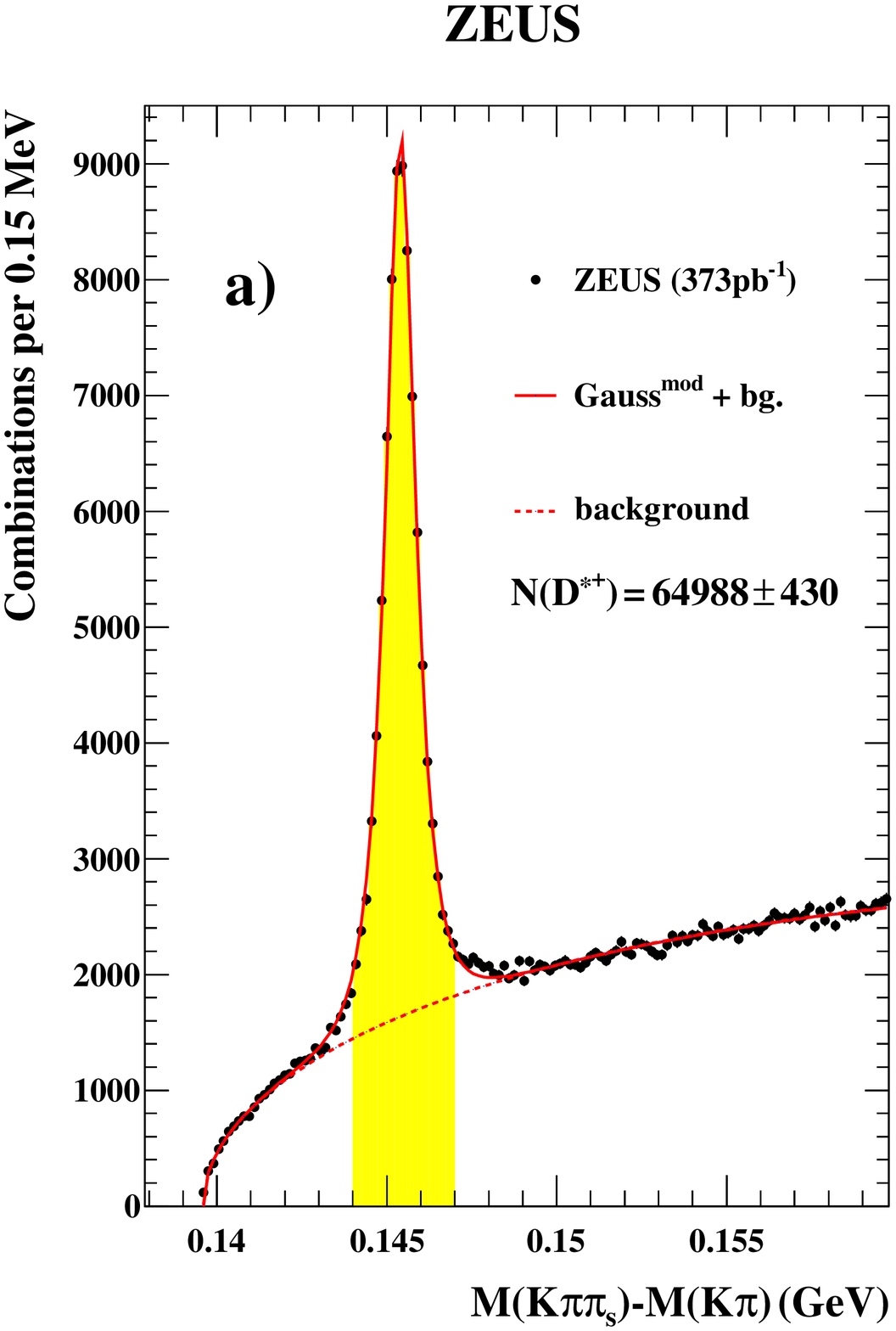}
\includegraphics[scale=0.40]{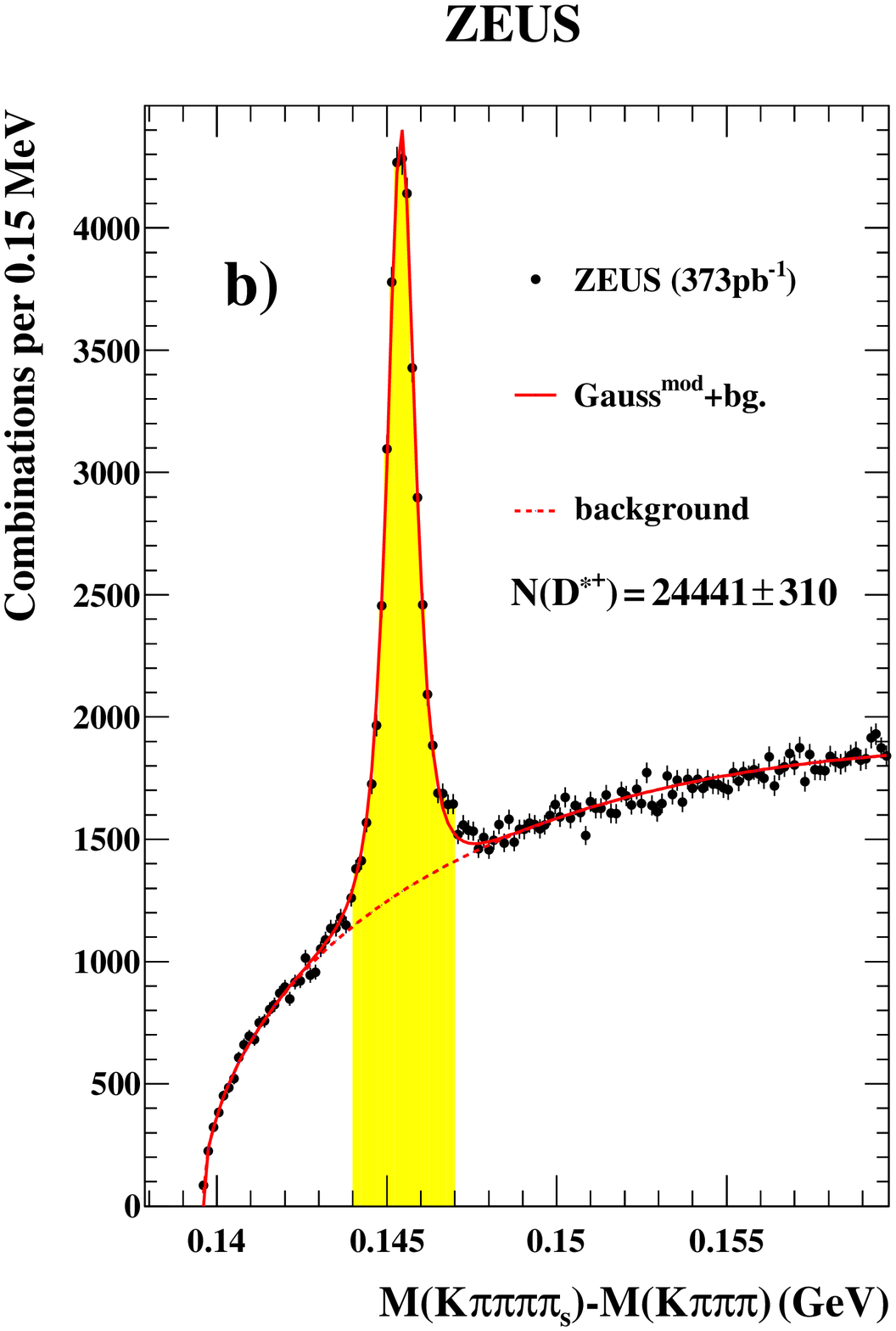}
\end{center}
\caption{The distribution of the mass difference (dots),
(a) $\Delta M = M(K\pi\pi_s) - M(K\pi)$ and 
(b) $\Delta M = M(K\pi\pi\pi\pi_s) - M(K\pi\pi\pi)$.  
The solid curves are  fits to the sum of a modified Gaussian function
and a background function (dashed lines). Candidates from the shaded area, $0.144 - 0.147\,\rm{GeV}$, are used for the
analysis of excited charm mesons.
        }
 
\label{1}
\end{figure}
} 

\newcommand{\figuretwo}{
\begin{figure}[ht]
\vfill
\begin{center}
\includegraphics[scale=0.40]{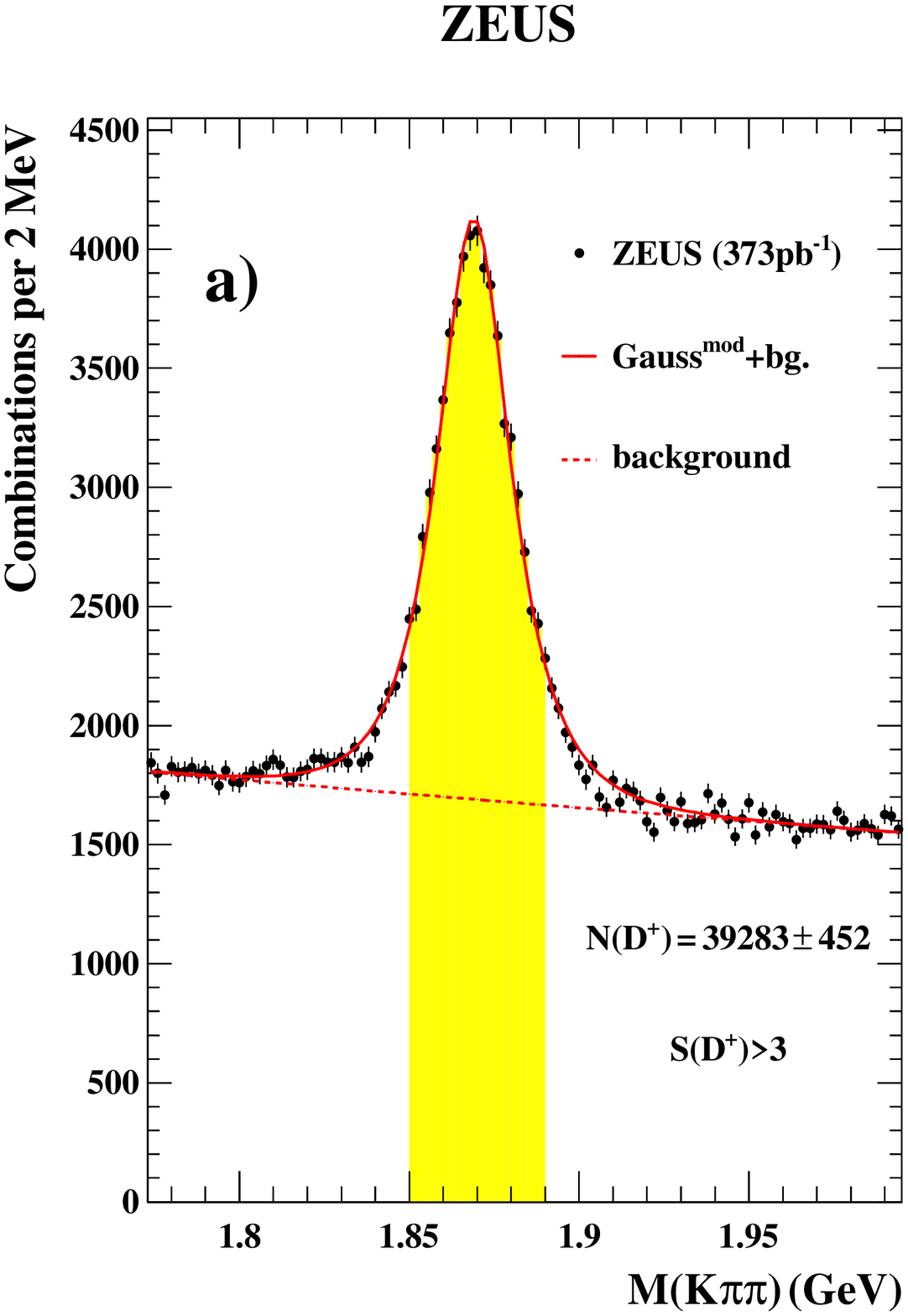}
\includegraphics[scale=0.40]{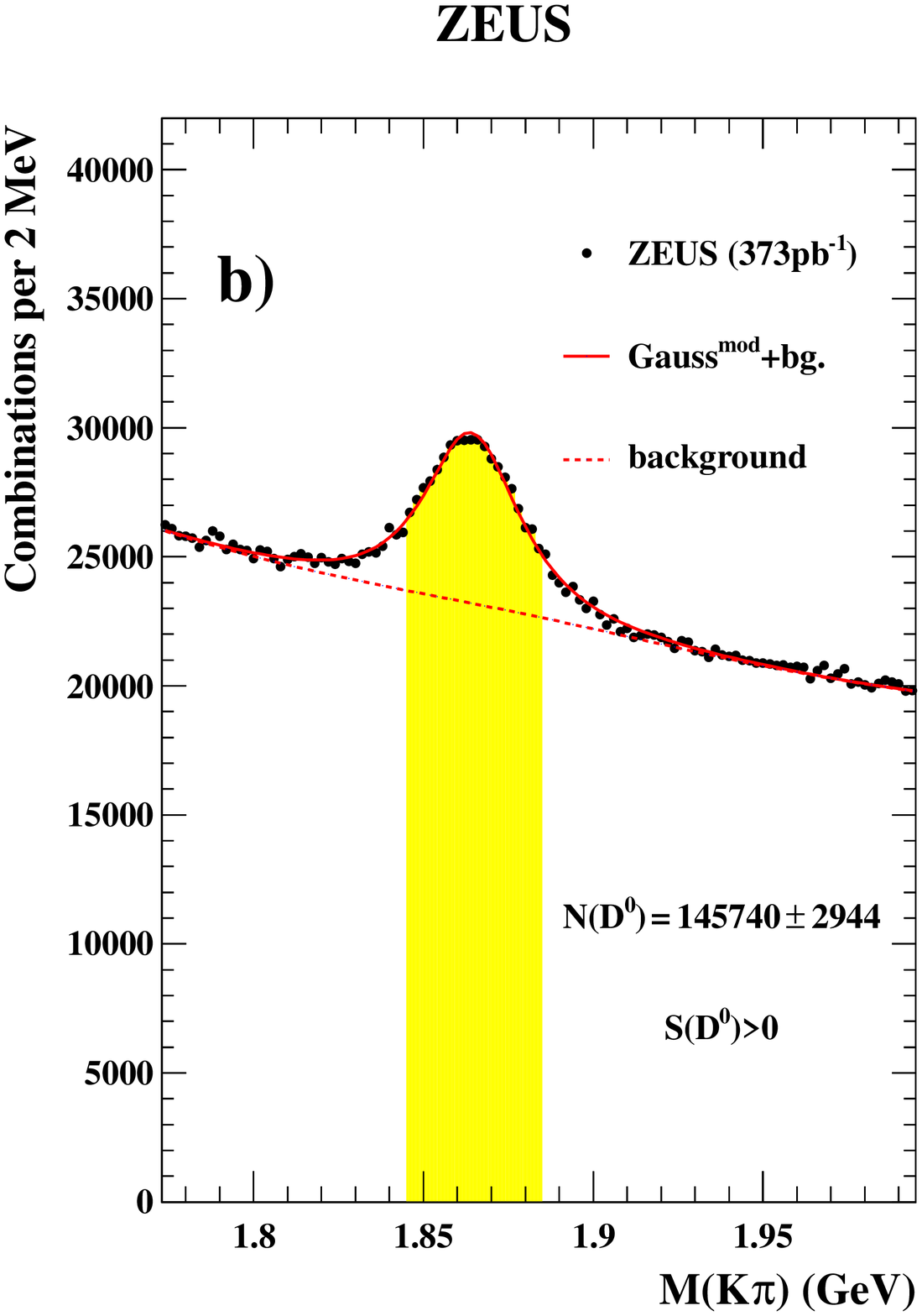}
\end{center}
\caption{The mass distributions (dots), (a) $M(K^-\pi^+\pi^+)$ for events with significance $S > 3$ and
                           (b) $M(K^-\pi^+)$ for events with significance $S > 0$.
The solid curves are fits to the sum of
   a modified Gaussian and a background function (dashed lines)
   and for (b) including also a contribution from a second broad
   modified Gaussian representing a reflection (see text). 
  Candidates from the shaded areas, (a) $1.85 - 1.89\,\rm{GeV}$ and (b) $1.845 - 1.885\,\rm{GeV}$,
are used for the analysis of excited
            charm mesons. }
 
\label{2}
\end{figure}
}
\newcommand{\figurethree}{
\begin{figure}[ht]
\vfill
\begin{center}
\includegraphics[scale=0.70]{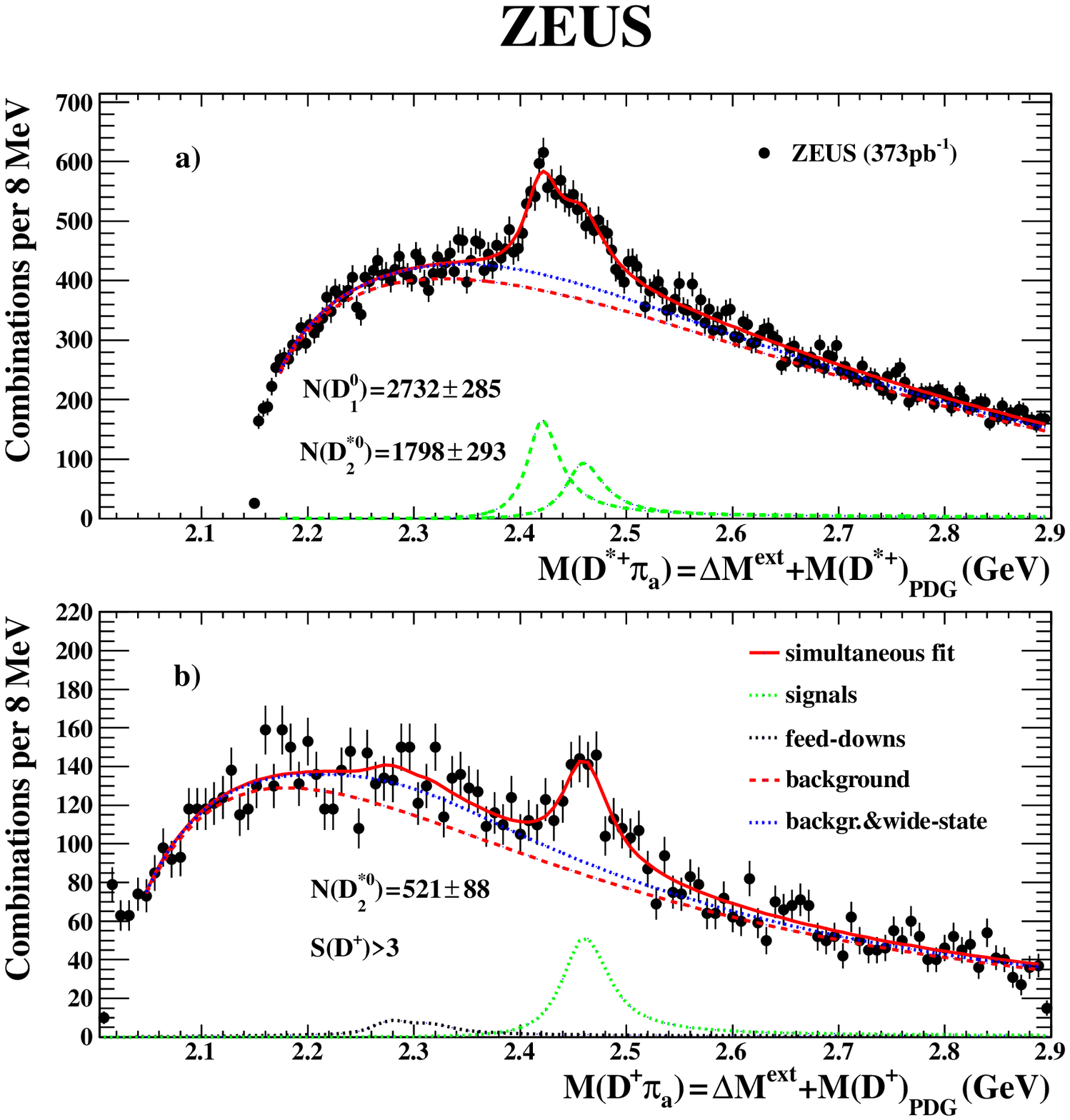}
\end{center}
\caption{                
The mass distributions (dots), a) $M(D^{*+}\pi_a)$  and b) $M(D^{+}\pi_a)$.
The solid curves are the result of a simultaneous fit to a) $D^0_1$ and $D^{*0}_2$  and
to b) $D^{*0}_2$ and feed-downs  plus background function (dashed curves).  The                 
contributions of the wide states $D_1(2430)^0$ and                                    
                      $D^*_0(2400)^0$ are given between the dashed and dotted curves.
The lowest curves are the contributions of the $D^0_1$, $D^{*0}_2$ and feed-downs to the fit.
        }  
\label{3}
\end{figure}
}
\newcommand{\figurefour}{
\begin{figure}[ht]
\vfill
\begin{center}
\includegraphics[scale=0.70]{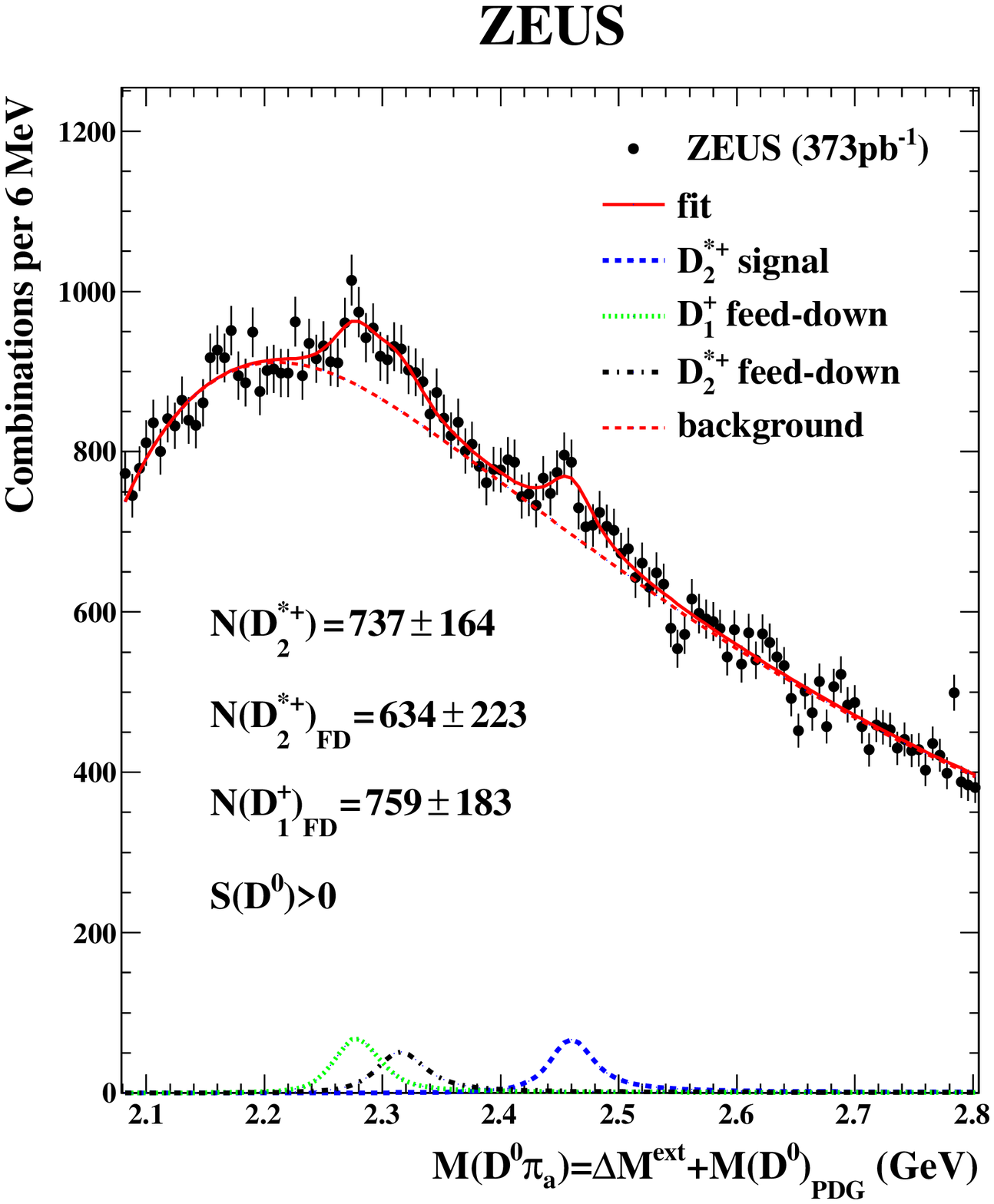}
\end{center}
\caption{                
The mass distribution (dots), $M(D^{0}\pi_a)$.
The solid curve is the result of a simultaneous fit to the feed-down (FD)           
 $D^+_1$ and $D^{*+}_2$ contributions and to the $D^{*+}_2$ signal
                           plus background function (dashed curves).                      
The lowest curves are the contributions of the $D^+_1$ and $D^{*+}_2$ to the fit.
        }  
\label{4}
\end{figure}
}
\newcommand{\figurefive}{
\begin{figure}[ht]
\vfill
\begin{center}
\includegraphics[scale=0.70]{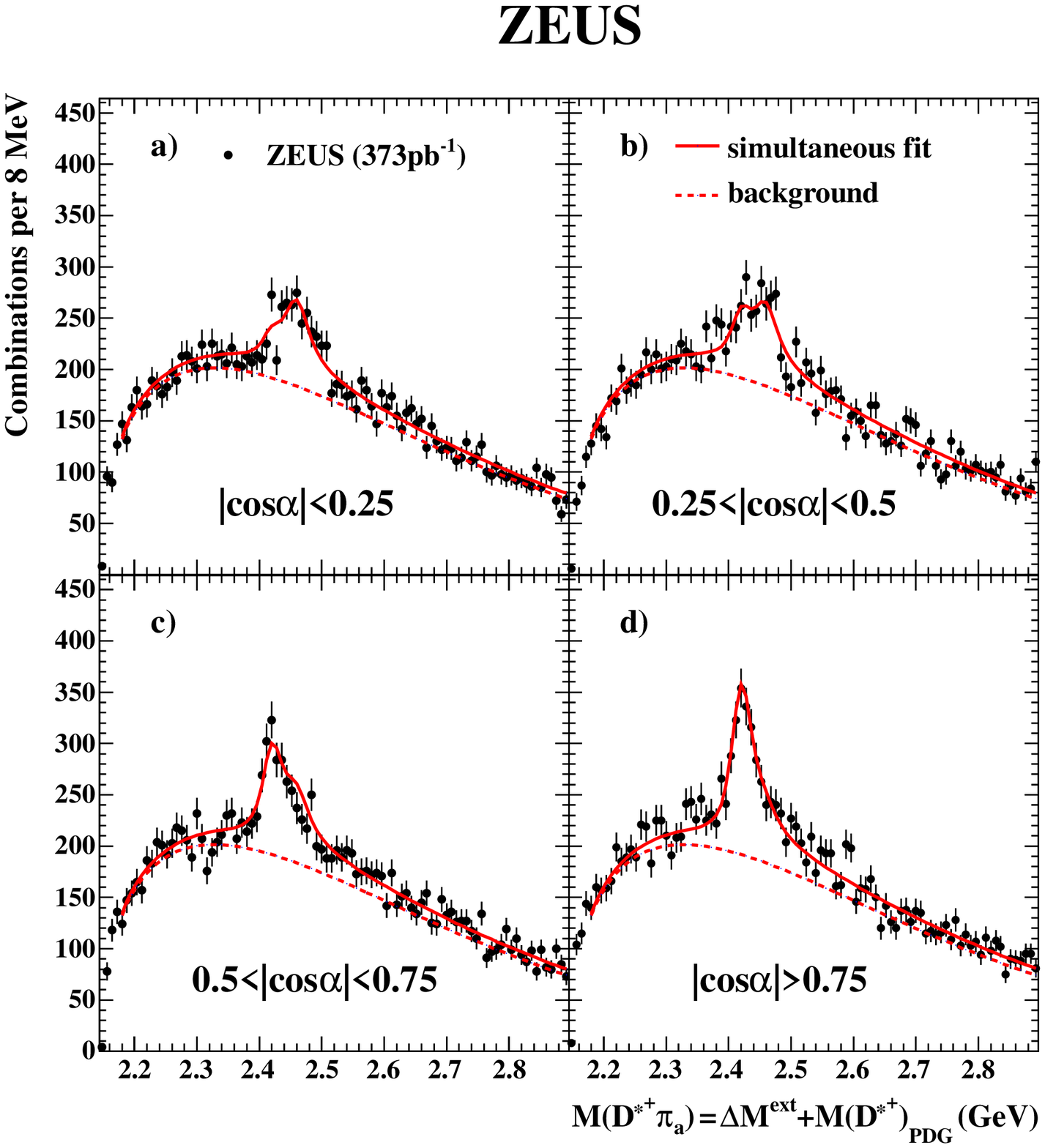}
\end{center}
\caption{                
The mass distributions (dots), $M(D^{*+}\pi_a)$ in four helicity intervals:  (a) $|\cos\alpha| < 0.25$;
(b) $0.25 < |\cos\alpha| < 0.50$; (c) $0.50 < |\cos\alpha| < 0.75$;  (d) $|\cos\alpha| > 0.75$.
 The       solid curves are the result of the simultaneous fit to
 $D^0_1$ and $D^{*0}_2$ plus background function (dashed curves).
         }
\label{5}
\end{figure}
}
\newcommand{\figuresix}{
\begin{figure}[ht]
\vfill
\begin{center}
\includegraphics[scale=0.70]{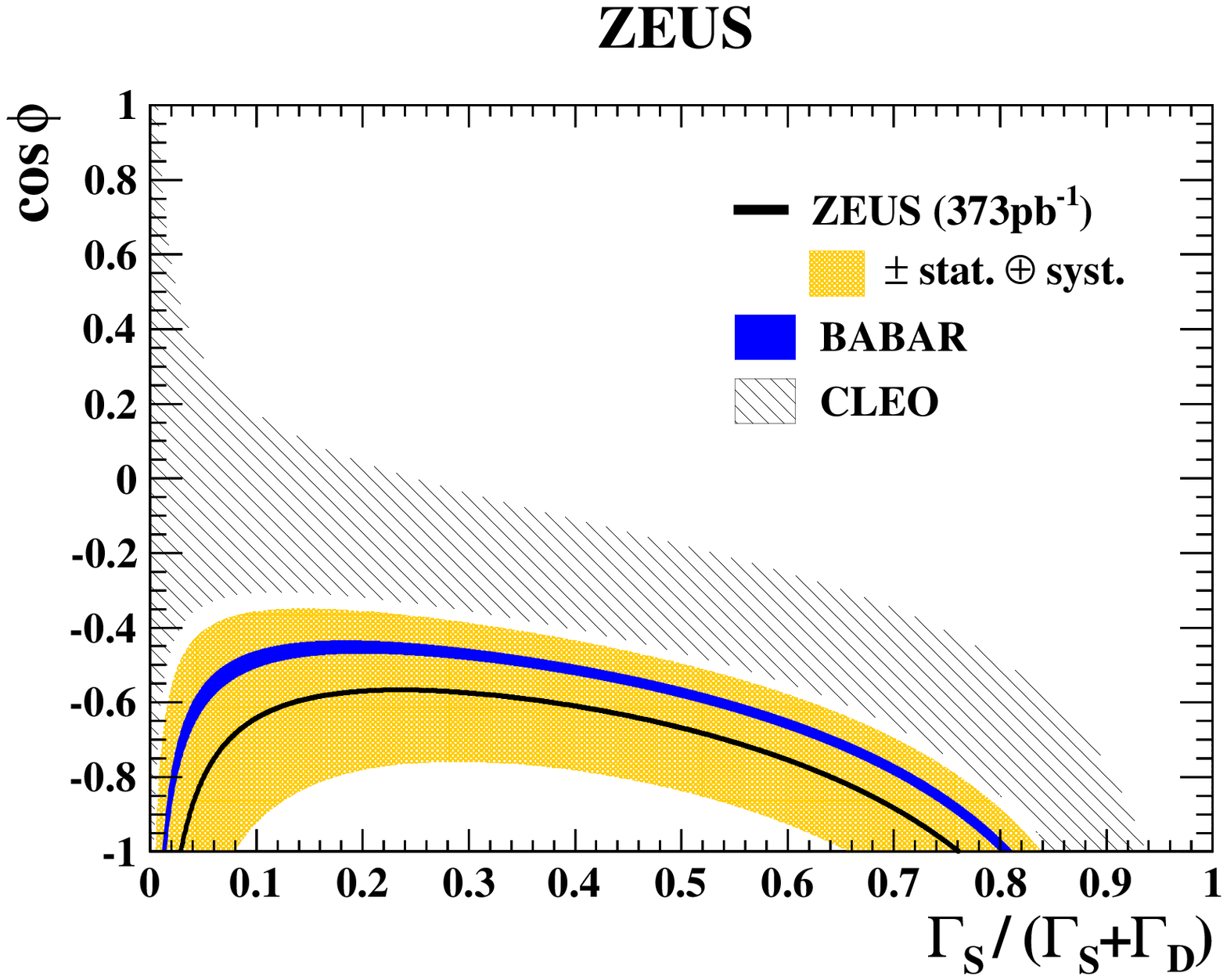}
\end{center}
\caption{  The allowed region of $\cos\phi$, where $\phi$ is
          the relative phase of $S$- and $D$-wave amplitudes,         
                     versus the fraction of $S$-wave
in the $D^0_1\to D^*\pi$ decay for ZEUS, BABAR and CLEO measurements.
         }
\label{6}
\end{figure}
\vfill
}
\begin{document}
\EDMtitle 
%
%
\pagenumbering{Roman} 

\begin{center}
{                      \Large  The ZEUS Collaboration              }
\end{center}

{\small


        {\raggedright
H.~Abramowicz$^{45, ah}$, 
I.~Abt$^{35}$, 
L.~Adamczyk$^{13}$, 
M.~Adamus$^{54}$, 
R.~Aggarwal$^{7, c}$, 
S.~Antonelli$^{4}$, 
P.~Antonioli$^{3}$, 
A.~Antonov$^{33}$, 
M.~Arneodo$^{50}$, 
O.~Arslan$^{5}$
% ...
        }

\newpage


\makebox[3em]{$^{1}$}
\begin{minipage}[t]{14cm}
{\it Argonne National Laboratory, Argonne, Illinois 60439-4815, USA}~$^{A}$

\end{minipage}\\
\makebox[3em]{$^{2}$}
\begin{minipage}[t]{14cm}
{\it Andrews University, Berrien Springs, Michigan 49104-0380, USA}

\end{minipage}\\
\makebox[3em]{$^{3}$}
\begin{minipage}[t]{14cm}
{\it INFN Bologna, Bologna, Italy}~$^{B}$

\end{minipage}\\
\makebox[3em]{$^{4}$}
\begin{minipage}[t]{14cm}
{\it University and INFN Bologna, Bologna, Italy}~$^{B}$

\end{minipage}\\
\makebox[3em]{$^{5}$}
\begin{minipage}[t]{14cm}
{\it Physikalisches Institut der Universit\"at Bonn,
Bonn, Germany}~$^{C}$

\end{minipage}\\
\makebox[3em]{$^{6}$}
\begin{minipage}[t]{14cm}
{\it H.H.~Wills Physics Laboratory, University of Bristol,
Bristol, United Kingdom}~$^{D}$

\end{minipage}\\
\makebox[3em]{$^{7}$}
\begin{minipage}[t]{14cm}
{\it Panjab University, Department of Physics, Chandigarh, India}

\end{minipage}\\
\makebox[3em]{$^{8}$}
\begin{minipage}[t]{14cm}
{\it Calabria University,
Physics Department and INFN, Cosenza, Italy}~$^{B}$

\end{minipage}\\
\makebox[3em]{$^{9}$}
\begin{minipage}[t]{14cm}
{\it Institute for Universe and Elementary Particles, Chonnam National University,\\
Kwangju, South Korea}

\end{minipage}\\
\makebox[3em]{$^{10}$}
\begin{minipage}[t]{14cm}
{\it Jabatan Fizik, Universiti Malaya, 50603 Kuala Lumpur, Malaysia}~$^{E}$

\end{minipage}\\
\makebox[3em]{$^{11}$}
\begin{minipage}[t]{14cm}
{\it Nevis Laboratories, Columbia University, Irvington on Hudson,
New York 10027, USA}~$^{F}$

\end{minipage}\\
\makebox[3em]{$^{12}$}
\begin{minipage}[t]{14cm}
{\it The Henryk Niewodniczanski Institute of Nuclear Physics, Polish Academy of \\
Sciences, Krakow, Poland}~$^{G}$

\end{minipage}\\
\makebox[3em]{$^{13}$}
\begin{minipage}[t]{14cm}
{\it AGH-University of Science and Technology, Faculty of Physics and Applied Computer
Science, Krakow, Poland}~$^{H}$

\end{minipage}\\
\makebox[3em]{$^{14}$}
\begin{minipage}[t]{14cm}
{\it Department of Physics, Jagellonian University, Cracow, Poland}

\end{minipage}\\
\makebox[3em]{$^{15}$}
\begin{minipage}[t]{14cm}
{\it Deutsches Elektronen-Synchrotron DESY, Hamburg, Germany}

\end{minipage}\\
\makebox[3em]{$^{16}$}
\begin{minipage}[t]{14cm}
{\it Deutsches Elektronen-Synchrotron DESY, Zeuthen, Germany}

\end{minipage}\\
\makebox[3em]{$^{17}$}
\begin{minipage}[t]{14cm}
{\it INFN Florence, Florence, Italy}~$^{B}$

\end{minipage}\\
\makebox[3em]{$^{18}$}
\begin{minipage}[t]{14cm}
{\it University and INFN Florence, Florence, Italy}~$^{B}$

\end{minipage}\\
\makebox[3em]{$^{19}$}
\begin{minipage}[t]{14cm}
{\it Fakult\"at f\"ur Physik der Universit\"at Freiburg i.Br.,
Freiburg i.Br., Germany}

\end{minipage}\\
\makebox[3em]{$^{20}$}
\begin{minipage}[t]{14cm}
{\it School of Physics and Astronomy, University of Glasgow,
Glasgow, United Kingdom}~$^{D}$

\end{minipage}\\
\makebox[3em]{$^{21}$}
\begin{minipage}[t]{14cm}
{\it Department of Engineering in Management and Finance, Univ. of
the Aegean, Chios, Greece}

\end{minipage}\\
\makebox[3em]{$^{22}$}
\begin{minipage}[t]{14cm}
{\it Hamburg University, Institute of Experimental Physics, Hamburg,
Germany}~$^{I}$

\end{minipage}\\
\makebox[3em]{$^{23}$}
\begin{minipage}[t]{14cm}
{\it Imperial College London, High Energy Nuclear Physics Group,
London, United Kingdom}~$^{D}$

\end{minipage}\\
\makebox[3em]{$^{24}$}
\begin{minipage}[t]{14cm}
{\it Institute of Particle and Nuclear Studies, KEK,
Tsukuba, Japan}~$^{J}$

\end{minipage}\\
\makebox[3em]{$^{25}$}
\begin{minipage}[t]{14cm}
{\it Institute of Physics and Technology of Ministry of Education and
Science of Kazakhstan, Almaty, Kazakhstan}

\end{minipage}\\
\makebox[3em]{$^{26}$}
\begin{minipage}[t]{14cm}
{\it Institute for Nuclear Research, National Academy of Sciences, Kyiv, Ukraine}

\end{minipage}\\
\makebox[3em]{$^{27}$}
\begin{minipage}[t]{14cm}
{\it Department of Nuclear Physics, National Taras Shevchenko University of Kyiv, Kyiv, Ukraine}

\end{minipage}\\
\makebox[3em]{$^{28}$}
\begin{minipage}[t]{14cm}
{\it Kyungpook National University, Center for High Energy Physics, Daegu,
South Korea}~$^{K}$

\end{minipage}\\
\makebox[3em]{$^{29}$}
\begin{minipage}[t]{14cm}
{\it Institut de Physique Nucl\'{e}aire, Universit\'{e} Catholique de Louvain, Louvain-la-Neuve,\\
Belgium}~$^{L}$

\end{minipage}\\
\makebox[3em]{$^{30}$}
\begin{minipage}[t]{14cm}
{\it Departamento de F\'{\i}sica Te\'orica, Universidad Aut\'onoma
de Madrid, Madrid, Spain}~$^{M}$

\end{minipage}\\
\makebox[3em]{$^{31}$}
\begin{minipage}[t]{14cm}
{\it Department of Physics, McGill University,
Montr\'eal, Qu\'ebec, Canada H3A 2T8}~$^{N}$

\end{minipage}\\
\makebox[3em]{$^{32}$}
\begin{minipage}[t]{14cm}
{\it Meiji Gakuin University, Faculty of General Education,
Yokohama, Japan}~$^{J}$

\end{minipage}\\
\makebox[3em]{$^{33}$}
\begin{minipage}[t]{14cm}
{\it Moscow Engineering Physics Institute, Moscow, Russia}~$^{O}$

\end{minipage}\\
\makebox[3em]{$^{34}$}
\begin{minipage}[t]{14cm}
{\it Lomonosov Moscow State University, Skobeltsyn Institute of Nuclear Physics,
Moscow, Russia}~$^{P}$

\end{minipage}\\
\makebox[3em]{$^{35}$}
\begin{minipage}[t]{14cm}
{\it Max-Planck-Institut f\"ur Physik, M\"unchen, Germany}

\end{minipage}\\
\makebox[3em]{$^{36}$}
\begin{minipage}[t]{14cm}
{\it NIKHEF and University of Amsterdam, Amsterdam, Netherlands}~$^{Q}$

\end{minipage}\\
\makebox[3em]{$^{37}$}
\begin{minipage}[t]{14cm}
{\it Physics Department, Ohio State University,
Columbus, Ohio 43210, USA}~$^{A}$

\end{minipage}\\
\makebox[3em]{$^{38}$}
\begin{minipage}[t]{14cm}
{\it Department of Physics, University of Oxford,
Oxford, United Kingdom}~$^{D}$

\end{minipage}\\
\makebox[3em]{$^{39}$}
\begin{minipage}[t]{14cm}
{\it INFN Padova, Padova, Italy}~$^{B}$

\end{minipage}\\
\makebox[3em]{$^{40}$}
\begin{minipage}[t]{14cm}
{\it Dipartimento di Fisica dell' Universit\`a and INFN,
Padova, Italy}~$^{B}$

\end{minipage}\\
\makebox[3em]{$^{41}$}
\begin{minipage}[t]{14cm}
{\it Department of Physics, Pennsylvania State University, University Park,\\
Pennsylvania 16802, USA}~$^{F}$

\end{minipage}\\
\makebox[3em]{$^{42}$}
\begin{minipage}[t]{14cm}
{\it Polytechnic University, Tokyo, Japan}~$^{J}$

\end{minipage}\\
\makebox[3em]{$^{43}$}
\begin{minipage}[t]{14cm}
{\it Dipartimento di Fisica, Universit\`a 'La Sapienza' and INFN,
Rome, Italy}~$^{B}$

\end{minipage}\\
\makebox[3em]{$^{44}$}
\begin{minipage}[t]{14cm}
{\it Rutherford Appleton Laboratory, Chilton, Didcot, Oxon,
United Kingdom}~$^{D}$

\end{minipage}\\
\makebox[3em]{$^{45}$}
\begin{minipage}[t]{14cm}
{\it Raymond and Beverly Sackler Faculty of Exact Sciences, School of Physics, \\
Tel Aviv University, Tel Aviv, Israel}~$^{R}$

\end{minipage}\\
\makebox[3em]{$^{46}$}
\begin{minipage}[t]{14cm}
{\it Department of Physics, Tokyo Institute of Technology,
Tokyo, Japan}~$^{J}$

\end{minipage}\\
\makebox[3em]{$^{47}$}
\begin{minipage}[t]{14cm}
{\it Department of Physics, University of Tokyo,
Tokyo, Japan}~$^{J}$

\end{minipage}\\
\makebox[3em]{$^{48}$}
\begin{minipage}[t]{14cm}
{\it Tokyo Metropolitan University, Department of Physics,
Tokyo, Japan}~$^{J}$

\end{minipage}\\
\makebox[3em]{$^{49}$}
\begin{minipage}[t]{14cm}
{\it Universit\`a di Torino and INFN, Torino, Italy}~$^{B}$

\end{minipage}\\
\makebox[3em]{$^{50}$}
\begin{minipage}[t]{14cm}
{\it Universit\`a del Piemonte Orientale, Novara, and INFN, Torino,
Italy}~$^{B}$

\end{minipage}\\
\makebox[3em]{$^{51}$}
\begin{minipage}[t]{14cm}
{\it Department of Physics, University of Toronto, Toronto, Ontario,
Canada M5S 1A7}~$^{N}$

\end{minipage}\\
\makebox[3em]{$^{52}$}
\begin{minipage}[t]{14cm}
{\it Physics and Astronomy Department, University College London,
London, United Kingdom}~$^{D}$

\end{minipage}\\
\makebox[3em]{$^{53}$}
\begin{minipage}[t]{14cm}
{\it Faculty of Physics, University of Warsaw, Warsaw, Poland}

\end{minipage}\\
\makebox[3em]{$^{54}$}
\begin{minipage}[t]{14cm}
{\it National Centre for Nuclear Research, Warsaw, Poland}

\end{minipage}\\
\makebox[3em]{$^{55}$}
\begin{minipage}[t]{14cm}
{\it Department of Particle Physics and Astrophysics, Weizmann
Institute, Rehovot, Israel}

\end{minipage}\\
\makebox[3em]{$^{56}$}
\begin{minipage}[t]{14cm}
{\it Department of Physics, University of Wisconsin, Madison,
Wisconsin 53706, USA}~$^{A}$

\end{minipage}\\
\makebox[3em]{$^{57}$}
\begin{minipage}[t]{14cm}
{\it Department of Physics, York University, Ontario, Canada M3J 1P3}~$^{N}$

\end{minipage}\\
\vspace{30em} \pagebreak[4]


\makebox[3ex]{$^{ A}$}
\begin{minipage}[t]{14cm}
 supported by the US Department of Energy\
\end{minipage}\\
\makebox[3ex]{$^{ B}$}
\begin{minipage}[t]{14cm}
 supported by the Italian National Institute for Nuclear Physics (INFN) \
\end{minipage}\\
\makebox[3ex]{$^{ C}$}
\begin{minipage}[t]{14cm}
 supported by the German Federal Ministry for Education and Research (BMBF), under
 contract No. 05 H09PDF\
\end{minipage}\\
\makebox[3ex]{$^{ D}$}
\begin{minipage}[t]{14cm}
 supported by the Science and Technology Facilities Council, UK\
\end{minipage}\\
\makebox[3ex]{$^{ E}$}
\begin{minipage}[t]{14cm}
 supported by an FRGS grant from the Malaysian government\
\end{minipage}\\
\makebox[3ex]{$^{ F}$}
\begin{minipage}[t]{14cm}
 supported by the US National Science Foundation. Any opinion,
 findings and conclusions or recommendations expressed in this material
 are those of the authors and do not necessarily reflect the views of the
 National Science Foundation.\
\end{minipage}\\
\makebox[3ex]{$^{ G}$}
\begin{minipage}[t]{14cm}
 supported by the Polish Ministry of Science and Higher Education as a scientific project No.
 DPN/N188/DESY/2009\
\end{minipage}\\
\makebox[3ex]{$^{ H}$}
\begin{minipage}[t]{14cm}
 supported by the Polish Ministry of Science and Higher Education and its grants
 for Scientific Research\
\end{minipage}\\
\makebox[3ex]{$^{ I}$}
\begin{minipage}[t]{14cm}
 supported by the German Federal Ministry for Education and Research (BMBF), under
 contract No. 05h09GUF, and the SFB 676 of the Deutsche Forschungsgemeinschaft (DFG) \
\end{minipage}\\
\makebox[3ex]{$^{ J}$}
\begin{minipage}[t]{14cm}
 supported by the Japanese Ministry of Education, Culture, Sports, Science and Technology
 (MEXT) and its grants for Scientific Research\
\end{minipage}\\
\makebox[3ex]{$^{ K}$}
\begin{minipage}[t]{14cm}
 supported by the Korean Ministry of Education and Korea Science and Engineering
 Foundation\
\end{minipage}\\
\makebox[3ex]{$^{ L}$}
\begin{minipage}[t]{14cm}
 supported by FNRS and its associated funds (IISN and FRIA) and by an Inter-University
 Attraction Poles Programme subsidised by the Belgian Federal Science Policy Office\
\end{minipage}\\
\makebox[3ex]{$^{ M}$}
\begin{minipage}[t]{14cm}
 supported by the Spanish Ministry of Education and Science through funds provided by
 CICYT\
\end{minipage}\\
\makebox[3ex]{$^{ N}$}
\begin{minipage}[t]{14cm}
 supported by the Natural Sciences and Engineering Research Council of Canada (NSERC) \
\end{minipage}\\
\makebox[3ex]{$^{ O}$}
\begin{minipage}[t]{14cm}
 partially supported by the German Federal Ministry for Education and Research (BMBF)\
\end{minipage}\\
\makebox[3ex]{$^{ P}$}
\begin{minipage}[t]{14cm}
 supported by RF Presidential grant N 4142.2010.2 for Leading Scientific Schools, by the
 Russian Ministry of Education and Science through its grant for Scientific Research on
 High Energy Physics and under contract No.02.740.11.0244 \
\end{minipage}\\
\makebox[3ex]{$^{ Q}$}
\begin{minipage}[t]{14cm}
 supported by the Netherlands Foundation for Research on Matter (FOM)\
\end{minipage}\\
\makebox[3ex]{$^{ R}$}
\begin{minipage}[t]{14cm}
 supported by the Israel Science Foundation\
\end{minipage}\\
\vspace{30em} \pagebreak[4]


\makebox[3ex]{$^{ a}$}
\begin{minipage}[t]{14cm}
now at University of Salerno, Italy\
\end{minipage}\\
\makebox[3ex]{$^{ b}$}
\begin{minipage}[t]{14cm}
now at Queen Mary University of London, United Kingdom\
\end{minipage}\\
\makebox[3ex]{$^{ c}$}
\begin{minipage}[t]{14cm}
also funded by Max Planck Institute for Physics, Munich, Germany\
\end{minipage}\\
\makebox[3ex]{$^{ d}$}
\begin{minipage}[t]{14cm}
also Senior Alexander von Humboldt Research Fellow at Hamburg University,
 Institute of Experimental Physics, Hamburg, Germany\
\end{minipage}\\
\makebox[3ex]{$^{ e}$}
\begin{minipage}[t]{14cm}
also at Cracow University of Technology, Faculty of Physics,
 Mathemathics and Applied Computer Science, Poland\
\end{minipage}\\
\makebox[3ex]{$^{ f}$}
\begin{minipage}[t]{14cm}
supported by the research grant No. 1 P03B 04529 (2005-2008)\
\end{minipage}\\
\makebox[3ex]{$^{ g}$}
\begin{minipage}[t]{14cm}
supported by the Polish National Science Centre, project No. DEC-2011/01/BST2/03643\
\end{minipage}\\
\makebox[3ex]{$^{ h}$}
\begin{minipage}[t]{14cm}
now at Rockefeller University, New York, NY
 10065, USA\
\end{minipage}\\
\makebox[3ex]{$^{ i}$}
\begin{minipage}[t]{14cm}
now at DESY group FS-CFEL-1\
\end{minipage}\\
\makebox[3ex]{$^{ j}$}
\begin{minipage}[t]{14cm}
now at Institute of High Energy Physics, Beijing, China\
\end{minipage}\\
\makebox[3ex]{$^{ k}$}
\begin{minipage}[t]{14cm}
now at DESY group FEB, Hamburg, Germany\
\end{minipage}\\
\makebox[3ex]{$^{ l}$}
\begin{minipage}[t]{14cm}
also at Moscow State University, Russia\
\end{minipage}\\
\makebox[3ex]{$^{ m}$}
\begin{minipage}[t]{14cm}
now at University of Liverpool, United Kingdom\
\end{minipage}\\
\makebox[3ex]{$^{ n}$}
\begin{minipage}[t]{14cm}
now at CERN, Geneva, Switzerland\
\end{minipage}\\
\makebox[3ex]{$^{ o}$}
\begin{minipage}[t]{14cm}
also affiliated with Universtiy College London, UK\
\end{minipage}\\
\makebox[3ex]{$^{ p}$}
\begin{minipage}[t]{14cm}
now at Goldman Sachs, London, UK\
\end{minipage}\\
\makebox[3ex]{$^{ q}$}
\begin{minipage}[t]{14cm}
also at Institute of Theoretical and Experimental Physics, Moscow, Russia\
\end{minipage}\\
\makebox[3ex]{$^{ r}$}
\begin{minipage}[t]{14cm}
also at FPACS, AGH-UST, Cracow, Poland\
\end{minipage}\\
\makebox[3ex]{$^{ s}$}
\begin{minipage}[t]{14cm}
partially supported by Warsaw University, Poland\
\end{minipage}\\
\makebox[3ex]{$^{ t}$}
\begin{minipage}[t]{14cm}
now at Istituto Nucleare di Fisica Nazionale (INFN), Pisa, Italy\
\end{minipage}\\
\makebox[3ex]{$^{ u}$}
\begin{minipage}[t]{14cm}
now at Haase Energie Technik AG, Neum\"unster, Germany\
\end{minipage}\\
\makebox[3ex]{$^{ v}$}
\begin{minipage}[t]{14cm}
now at Department of Physics, University of Bonn, Germany\
\end{minipage}\\
\makebox[3ex]{$^{ w}$}
\begin{minipage}[t]{14cm}
also affiliated with DESY, Germany\
\end{minipage}\\
\makebox[3ex]{$^{ x}$}
\begin{minipage}[t]{14cm}
also at University of Tokyo, Japan\
\end{minipage}\\
\makebox[3ex]{$^{ y}$}
\begin{minipage}[t]{14cm}
now at Kobe University, Japan\
\end{minipage}\\
\makebox[3ex]{$^{ z}$}
\begin{minipage}[t]{14cm}
supported by DESY, Germany\
\end{minipage}\\
\makebox[3ex]{$^{\dagger}$}
\begin{minipage}[t]{14cm}
 deceased \
\end{minipage}\\
\makebox[3ex]{$^{aa}$}
\begin{minipage}[t]{14cm}
member of National Technical University of Ukraine, Kyiv Polytechnic Institute,
 Kyiv, Ukraine\
\end{minipage}\\
\makebox[3ex]{$^{ab}$}
\begin{minipage}[t]{14cm}
member of National University of Kyiv - Mohyla Academy, Kyiv, Ukraine\
\end{minipage}\\
\makebox[3ex]{$^{ac}$}
\begin{minipage}[t]{14cm}
partly supported by the Russian Foundation for Basic Research, grant 11-02-91345-DFG\_a\
\end{minipage}\\
\makebox[3ex]{$^{ad}$}
\begin{minipage}[t]{14cm}
Alexander von Humboldt Professor; also at DESY and University of Oxford\
\end{minipage}\\
\makebox[3ex]{$^{ae}$}
\begin{minipage}[t]{14cm}
STFC Advanced Fellow\
\end{minipage}\\
\makebox[3ex]{$^{af}$}
\begin{minipage}[t]{14cm}
now at LNF, Frascati, Italy\
\end{minipage}\\
\makebox[3ex]{$^{ag}$}
\begin{minipage}[t]{14cm}
This material was based on work supported by the
 National Science Foundation, while working at the Foundation.\
\end{minipage}\\
\makebox[3ex]{$^{ah}$}
\begin{minipage}[t]{14cm}
also at Max Planck Institute for Physics, Munich, Germany, External Scientific Member\
\end{minipage}\\
\makebox[3ex]{$^{ai}$}
\begin{minipage}[t]{14cm}
now at Tokyo Metropolitan University, Japan\
\end{minipage}\\
\makebox[3ex]{$^{aj}$}
\begin{minipage}[t]{14cm}
now at Nihon Institute of Medical Science, Japan\
\end{minipage}\\
\makebox[3ex]{$^{ak}$}
\begin{minipage}[t]{14cm}
now at Osaka University, Osaka, Japan\
\end{minipage}\\
\makebox[3ex]{$^{al}$}
\begin{minipage}[t]{14cm}
also at \L\'{o}d\'{z} University, Poland\
\end{minipage}\\
\makebox[3ex]{$^{am}$}
\begin{minipage}[t]{14cm}
member of \L\'{o}d\'{z} University, Poland\
\end{minipage}\\
\makebox[3ex]{$^{an}$}
\begin{minipage}[t]{14cm}
now at Department of Physics, Stockholm University, Stockholm, Sweden\
\end{minipage}\\
\makebox[3ex]{$^{ao}$}
\begin{minipage}[t]{14cm}
also at Cardinal Stefan Wyszy\'nski University, Warsaw, Poland\
\end{minipage}\\

} 
\clearpage
\pagenumbering{arabic}
\pagestyle{plain}
\section{Introduction}
\label{sec-int}
 
The production of the well-established ground-state charm
 mesons $D$ and $D^*$ has been extensively studied in $ep$ collisions at HERA. 
The large charm production cross section at HERA makes it possible
 to also investigate the excited charm-meson states.
In a previous ZEUS analysis~\cite{dsshera1},        
        with an integrated luminosity of $126\pbi$,              
                                                              the orbitally excited states
$D_1(2420)^0$ with $J^P = 1^+$ and $D^*_2(2460)^0$ with $J^P = 2^+$ were studied
in the decay modes\footnote{The corresponding anti-particle decays were also measured.                          
                            Hereafter, charge conjugation is implied.}                                      
                    $D_1(2420)^0\to D^{*}(2010)^{+}\pi^{-}$ and 
                    $D^*_2(2460)^0\to D^{*}(2010)^{+}\pi^{-}$, $D^{+}\pi^{-}$.
The width of the $D^0_1$ 
was   found to be        
significantly above the 2008 world-average value~\cite{PDG}.
A study of the helicity angular distribution of the $D_1(2420)^0$ gave results that were consistent with some 
$S$-wave admixture in the decay $D_1^0\to D^{*+}\pi^{-}$, 
contrary to Heavy Quark Effective Theory (HQET) predictions~\cite{HQET1,HQET2} and to previous experimental results~\cite{Belle} 
 which had yielded a pure $D$-wave decay in this channel.
 
In this paper the analysis was repeated with an independent data sample of higher integrated
luminosity. In addition the production of the charged excited charm mesons                              
$D_1(2420)^{+}$ and $D^*_2(2460)^{+}$ was studied for the first time at HERA
in the decay modes $D_1(2420)^{+}\to D^{*}(2007)^{0}\pi^{+}$ and 
                    $D^*_2(2460)^{+}\to D^{*}(2007)^{0}\pi^{+}$, $D^{0}\pi^{+}$.
For both the neutral and charged excited charm mesons the study also includes 
a measurement of fragmentation fractions and ratios of the $D^*_2$ branching fractions.   
 
The analysis was performed using data taken from 2003 to 2007, when HERA collided electrons
or positrons at $27.5$~GeV with protons at $920$~GeV. The data correspond to an integrated
luminosity of $373$ pb$^{-1}$.
            The upgraded ZEUS detector        included 
a microvertex detector, allowing the measurement of the decay vertex of charm mesons. In particular,
the signal-to-background ratio was significantly improved for the $D^{+}$ meson,
                                                  which has the highest lifetime
among the charm hadrons.
 
To maximise the statistics,
 both photoproduction and deep inelastic scattering  events were used in this analysis.
Events produced in the photoproduction regime contributed $70 - 80\%$ of the selected charm-meson samples.

\section{Experimental set-up}
\label{sec-exp}

A detailed description of the ZEUS detector can be found 
elsewhere~\cite{zeus:1993:bluebook}. A brief outline of the 
components that are most relevant for this analysis is given
below.\xspace
 
 
In the kinematic range of the analysis, charged particles were tracked
in the central tracking detector (CTD)~\cite{nim:a279:290,*npps:b32:181,*nim:a338:254} and the microvertex
detector (MVD)~\cite{nim:a581:656}. These components operated in a magnetic
field of \unit{1.43}{\tesla} provided by a thin superconducting solenoid. The
CTD consisted of 72~cylindrical drift-chamber layers, organised in nine
superlayers covering the polar-angle\footnote{The ZEUS coordinate system is a right-handed Cartesian system, with the $Z$ 
axis pointing in the nominal proton beam direction, referred to as the ``forward
direction'', and the $X$ axis pointing left towards the centre of HERA.
The coordinate origin is at the centre of the CTD.\xspace The pseudorapidity is defined as $\eta=-\ln\left(\tan\frac{\theta}{2}\right)$,
where the polar angle, $\theta$, is measured with respect to the
$Z$ axis.\xspace}~region
\mbox{$\unit{15}{\degree}<\theta<\unit{164}{\degree}$}.
The MVD silicon tracker consisted of a barrel (BMVD) and a forward
(FMVD) section. The BMVD contained three layers and provided
polar-angle coverage for tracks from $\unit{30}{\degree}$ to
$\unit{150}{\degree}$. The four-layer FMVD extended the polar-angle coverage in
the forward region to $\unit{7}{\degree}$. After alignment, the single-hit
resolution of the MVD was \unit{24}{\micron}. The transverse distance of closest
approach (DCA) of tracks  to the nominal vertex in the $X$--$Y$ plane was measured to have
a resolution, averaged over the azimuthal angle, of \unit{$(46 \oplus 122 /
p_{T})$}{\micron}, with $p_{T}$ in \GeV.  For CTD-MVD tracks that pass
through all nine CTD superlayers, the momentum resolution was
$\sigma(p_{T})/p_{T} = 0.0029 p_{T} \oplus 0.0081 \oplus
0.0012/p_{T}$, with $p_{T}$ in \GeV.
 
The high-resolution uranium--scintillator calorimeter (CAL)~\cite{nim:a309:77,*nim:a309:101,*nim:a321:356,*nim:a336:23}
consisted of three parts: the forward (FCAL), the barrel (BCAL) and
the rear (RCAL) calorimeters. Each part was subdivided transversely
into towers and longitudinally into one electromagnetic section (EMC)
and either one (in RCAL) or two (in BCAL and FCAL) hadronic sections
(HAC). The smallest subdivision of the calorimeter was called a cell.
The CAL energy resolutions, as measured under test-beam conditions,
were $\sigma(E)/E=0.18/\sqrt{E}$ for electrons and
$\sigma(E)/E=0.35/\sqrt{E}$ for hadrons, with $E$ in \GeV.
 
The luminosity was measured using the Bethe-Heitler reaction
$ep\,\rightarrow\, e\gamma p$ by a detector which consisted
of an independent lead--scintillator calorimeter~\cite{
desy-92-066,*zfp:c63:391,*acpp:b32:2025} and a magnetic
spectrometer\cite{nim:a565:572} system. 
 
\section{Event simulation}
\label{sec-sim}

Monte Carlo (MC) samples of charm and beauty events
were produced with the
 {\sc Pythia} 6.221\cite{pythia6221} and the
{\sc Rapgap}~3.000\cite{rapgap3}
event generators.
The generation included direct photon processes,
in which the photon couples directly to a parton in the proton,
and resolved photon processes, where the photon acts as a source
of partons, one of which participates in the hard scattering process.
The CTEQ5L~\cite{epj:c12:375} and the GRV~LO~\cite{pr:d46:1973} parametrisations
were used for the proton and photon parton density functions, respectively.
The charm- and beauty-quark masses were set to $1.5\,$GeV and $4.75\,$GeV,
respectively.
The  masses and widths for charm mesons were set to the latest PDG~\cite{PDG11} values. 
 
Events for all processes were generated in proportion to
the respective MC cross sections.
The Lund string model
was used for hadronisation in {\sc Pythia} and {\sc Rapgap}.
The Bowler modification~\cite{zfp:c11:169}
of the Lund symmetric fragmentation function~\cite{zfp:c20:317}
was used for the charm- and beauty-quark fragmentation.

The {\sc Pythia} and {\sc Rapgap} generators
were tuned to describe the photoproduction and the deep inelastic scattering
 regimes,
respectively~\cite{dsshera1}.
Subsequently,
the {\sc Pythia} events, generated with $Q^2<1.5\,\mbox{GeV}^2$,
were combined with the {\sc Rapgap} events, generated with $Q^2>1.5\,\mbox{GeV}^2$,
where $Q^2$ is the exchanged-photon virtuality.
 
The generated events were passed through a full simulation
of the detector using {\sc Geant} 3.13~\cite{geant}
and processed with the same reconstruction program as used for the data.
 
\section{Event selection and reconstruction of ground-state charm mesons }
\label{sec-sel}
 
The ZEUS trigger chain had three levels~\cite{zeus:1993:bluebook,Smith:1992im,trigger}. The first- and second-level trigger used
CAL and CTD data to select $ep$ collisions and to reject 
beam-gas events.
At the third-level trigger, 
the full event information was available. All relevant trigger chains were used for the data.
Triggers that required the presence of a reconstructed $D^{*+}\to D^0\pi^+\to (K^-\pi^+)\pi^+$ or
 $(K^-\pi^+\pi^-\pi^+)\pi^+$, $D^+\to K^-\pi^+\pi^+$ or $D^0\to K^-\pi^+$ candidate constituted a
 major fraction of the selected events. However, events missed by these triggers but selected with
 other trigger branches were also used in the analysis.
Applying, in the MC, either no trigger selection cuts or requiring at least 
one trigger chain to be passed did not affect the final measurements.
 
To ensure  high purity in the event sample, the $Z$ position of the primary vertex, reconstructed from CTD and MVD tracks, 
had to be within $|Z_{\mathrm{vtx}}| < 30$~cm.
All             charm mesons were reconstructed with tracks measured
in the CTD and MVD. 
All                 tracks were required to have a transverse momentum, $p_T$, above                             
               0.1~GeV, to start not further out than
the first CTD superlayer and to reach at least the third superlayer.
 The tracks were assigned either to the reconstructed primary vertex or to a secondary decay vertex
 associated with the weak   decay of a  charm meson, $D^{+}$ or $D^0$. To ensure
          the use of well reconstructed MVD tracks, all tracks associated  with the secondary vertex were required to                            
                    have at least two BMVD measurements in the $X$--$Y$ plane and two in the $Z$ direction.

The
    decay-length significance is a powerful tool for rejection of combinatorial background. It is defined as
                                          $S   = l/\sigma_l$,                               
 where the decay length $l$ is the distance in the transverse plane between the production
point and the decay vertex of a candidate charm     meson projected on its momentum direction and $\sigma_l$ is the 
uncertainty of this quantity~\cite{Dpaper}. The  quantity $S$ is positive when     the angle between the particle momenta and
the direction from primary to secondary vertex is less than $\pi/2$; it is negative otherwise.             
The $S$ distribution is asymmetric around zero,
                                                      with a stronger positive contribution coming mostly from
the charm mesons. The contributions to negative    $S$ values are due to background and resolution effects.
 
The combinatorial background was suppressed 
by selecting events above a minimum value of the ratio
$p_T (D)/E_{\perp}^{\theta > 10^{\circ}}$,                                       
where $D$ denotes $D^{*+},D^{+}$ or $D^0$ 
and
$E_{\perp}^{\theta > 10^{\circ}}$
is the transverse energy measured using all CAL cells outside a cone of $10^{\circ}$ around the
forward direction.
In addition, to reduce background, the 
$dE/dx$ values measured in the  CTD  of track candidates  originating from
the $D$ mesons 
were used. 
The parametrisation of the $dE/dx$ expectation values and the $\chi^2$ probabilities $l_K$ and $l_{\pi}$
 of the kaon and pion hypotheses, respectively,  were obtained as described                      
  in previous analyses~\cite{epj:c38:29,epj:c44:351}.
 The cuts $l_K > 0.03$ and $l_{\pi} > 0.01$ were applied.
 
\subsection{$\boldsymbol{D^{*+}}$ reconstruction}
\label{sec-dstar}
 
$D^{*+}$ mesons were identified via the decay modes
$D^{*+}\to D^0\pi^+_s\to (K^-\pi^+)\pi^+_s$ and                                    
$D^{*+}\to D^0\pi^+_s\to (K^-\pi^+\pi^-\pi^+)\pi^+_s$, where $\pi_s$ is a low-momentum       
     (``soft'') pion due to
the small mass difference  between $D^{*+}$ and $D^0$.                       
Tracks were combined to form $D^0$ candidates by calculating the invariant-mass   combinations
$M(K\pi)$ or $M(K\pi\pi\pi)$ with total charge zero. $D^{*+}$ candidates were formed by adding
a soft pion, $\pi_s$, with opposite charge to that of the kaon. 
Combinatorial background was reduced by
applying cuts as detailed in Table~\ref{tab1}.
 
The mass differences $\Delta M = M(K\pi\pi_s) - M(K\pi)$ and 
$\Delta M = M(K\pi\pi\pi\pi_s) - M(K\pi\pi\pi)$
were calculated for the $D^{*+}$ candidates that passed the cuts of Table~\ref{tab1}.
      Figure~\ref{1} shows the $\Delta M$ distributions for these
    $D^{*+}$ candidates. Clean peaks are seen at     the nominal value of $M(D^{*+})-M(D^0)$~\cite{PDG11}.
 
The $\Delta M$ distributions were fitted to a sum of a background function and                              
                                             a modified Gaussian function~\cite{dsshera1}.
The fit yielded 
 $D^{*+}$ signals  of \DStarTwoProngYield candidates for $D^0\to K\pi$ and
                         \DStarFourProngYield   candidates for $D^0\to K\pi\pi\pi$.
The fitted mass differences were \DStarTwoProngMass and \DStarFourProngMass respectively, in agreement with the PDG average value~\cite{PDG11}.
Only
$D^{*+}$ candidates with $0.144 < \Delta M < 0.147$~GeV were used for the excited charm mesons
analysis.
 
\subsection{$\boldsymbol{D^{+}}$ reconstruction}
\label{sec-dplus}

$D^{+}$ mesons were reconstructed from the decay $D^+\to K^-\pi^+\pi^+$  with looser     
kinematic  cuts
than in the previous analysis~\cite{dsshera1}, made possible by the cleaner identification with the MVD.
For each event, track pairs with equal charge and pion mass assignment were combined with
a track with opposite charge with a kaon mass assignment to form a $D^+$ candidate.
These tracks were refitted to a common decay vertex, and the invariant mass,
$M(K\pi\pi)$, was calculated. The $K$ and $\pi$ tracks were required to have transverse momentum
$p_T^K > 0.5$~GeV and $p_T^\pi > 0.35$~GeV and the distance of closest approach between each pair
of the three tracks was required to be less than $0.3$~cm.
To suppress combinatorial background, the following cuts were applied:   
 \begin{itemize}
\item  $\cos\theta^*(K) > -0.75$, where
$\theta^*(K)$ is the angle between the kaon in the $K\pi\pi$ rest frame and the $K\pi\pi$ line of
flight in the laboratory frame;  
\item  the $\chi^2$ of the fit of  the  decay vertex was  
less than 10;   
\item  the decay-length significance, $S(D^+)$, was greater than 3.             
 \end{itemize} 
Background from $D^{*+}$ decays was removed by requiring $M(K\pi\pi) - M(K\pi) > 0.15$~GeV.
Background from $D^+_s\to\phi\pi,\  \phi\to K^+K^-$ was suppressed by requiring
that  the invariant mass of any two $D^+$ decay candidate tracks with opposite charge should be 
outside
$\pm 8$~MeV around the nominal $\phi$ mass when the kaon mass was assigned to both tracks.
$D^+$ candidates in the kinematic range $p_T(D^+) > 2.8$~GeV and $|\eta(D^+)| < 1.6$ were kept
for further analysis.
  
Figure~\ref{2}~(a) shows the $M(K^-\pi^+\pi^+)$ distribution for $D^+$ candidates after the cuts.                    
A clear signal is
seen at the nominal value of the $D^{+}$ mass~\cite{PDG11}.
The mass distribution was fitted to a sum of a modified Gaussian function  and a
polynomial background. The fit yielded 
 a $D^{+}$ signal of \DChargeYield events
and a $D^{+}$ mass of \DChargeMass, in agreement with the PDG average value~\cite{PDG11}.
  Only
$D^{+}$ candidates with $1.85 < M(K\pi\pi) < 1.89$~GeV were used for the excited charm mesons
 analysis.
 
\subsection{$\boldsymbol{D^{0}}$ reconstruction}
\label{sec-dzero}

$D^{0}$ mesons were reconstructed from the decay $D^0\to K^-\pi^+$.                 
For each event, two tracks with opposite charge and $K$ and $\pi$ mass assignments, respectively,
                                                were combined to form a $D^0$ candidate.
These tracks were  refitted to a common decay vertex, and the invariant mass,
$M(K\pi)$, was calculated. Both tracks were required to have transverse momentum
$p_T^K > 0.5$~GeV and 
$p_T^{\pi} > 0.7$~GeV                           
                                           and the distance of closest approach between these   
             tracks was required to be less than $0.1$~cm.
To suppress combinatorial background, the following cuts were applied:   
 \begin{itemize}
\item  $|\cos\theta^*(K)| < 0.85$, where
$\theta^*(K)$ is the angle between the kaon in the $K\pi$ rest frame and the $K\pi$ line of
flight in the laboratory frame;   
      \item   the $\chi^2$ of the decay vertex was   
 less than 20; 
\item  the  decay-length significance, $S(D^0)$, was bigger than 0.           
 \end{itemize}
$D^0$ candidates in the kinematic range $p_T(D^0) > 2.6$~GeV and $|\eta(D^0)| < 1.6$ were kept
for further analysis.
  
Figure~\ref{2}~(b) shows the $M(K^-\pi^+)$ distribution for $D^0$ candidates after the cuts.                   
A clear signal is
seen at the nominal value of the $D^{0}$ mass~\cite{PDG11}.
The mass distribution was fitted to a sum of a modified Gaussian function,
a broad modified Gaussian representing the reflection produced 
by $D^0$ mesons with the wrong (opposite) kaon and pion mass assignment
and a polynomial background.
For the reflection, the shape parameters of the broad modified Gaussian were obtained
from a study of the MC signal sample and the normalisation (integral) was set equal to that
of the other modified Gaussian.
The fit yielded 
a $D^{0}$ signal of \DZeroYield events
and a $D^{0}$ mass of \DZeroMass which is $0.8$~MeV lower than the PDG average value~\cite{PDG11}.
This deviation does not affect any of the results of the excited charm mesons.
Only
$D^{0}$ candidates with $1.845 < M(K\pi\pi) < 1.885$~GeV were used for the excited charm mesons
 analysis.
 
\section{ $\boldsymbol{D_1}$ and $\boldsymbol{D^{*}_2}$ reconstruction}
\label{sec-dss}
          
\subsection{Reconstruction of the $\boldsymbol{D_1(2420)^0}$ and $\boldsymbol{D^*_2(2460)^0}$ mesons}
 \label{sec-dssone}
The $D_1(2420)^0$ and $D^*_2(2460)^0$ mesons were reconstructed in the decay mode 
 $D^{*+}\pi^-$ by combining each $D^{*+}$ candidate 
                                                             with an additional track, assumed to be
a pion ($\pi_a$), with a charge opposite to that of the $D^*$.
Combinatorial background was reduced by applying the following cuts:   
 \begin{itemize}
\item  $p_T (\pi_a) > 0.15$~GeV; 
\item  $\eta(\pi_a) < 1.1$ ;
\item  $p_T (D^{*+}\pi_a)/E_{\perp}^{\theta > 10^{\circ}} > 0.25$
($0.30$) for the $D^0\to K\pi$   
                                                                    ($D^0\to K\pi\pi\pi$) channel; 
\item  $\cos\theta^*(D^{*+}) < 0.9$, where
$\theta^*(D^{*+})$ is the angle between the $D^{*+}$ in the $D^{*+}\pi_a$ rest frame and the $D^{*+}\pi_a$ line of
flight in the laboratory frame; 
\item  the cut $l_{\pi} > 0.01$ was applied for $\pi_a$.   
 \end{itemize}
For each excited charm-meson candidate, the ``extended'' mass difference,
$\Delta M^{\rm{ext}} = M(K\pi\pi_s\pi_a) - M(K\pi\pi_s)$ or                                       
$\Delta M^{\rm{ext}} = M(K\pi\pi\pi\pi_s\pi_a) - M(K\pi\pi\pi\pi_s)$, was calculated.                          
                              Figure~\ref{3}~(a) shows the invariant mass 
$M(D^{*+}\pi_a) = \Delta M^{\rm{ext}} + M(D^{*+}_{\mathrm{PDG}})$, where $M(D^{*+}_{\mathrm{PDG}})$ 
is the nominal $D^{*+}$
mass~\cite{PDG11}. A clear signal in the $D^0_1/D^{*0}_2$ mass region is seen. 
 
The $D^*_2(2460)^0$ was also reconstructed  in the decay mode $D^*_2(2460)^0\to D^{+}\pi^-$ by combining each $D^+$    
candidate                                                
                with an additional track, assumed to be a pion $\pi_a$, with a charge opposite to that of the $D^+$.
Combinatorial background was reduced by applying the following cuts:    
 \begin{itemize}
\item  $p_T (\pi_a) > 0.3$~GeV;  
\item  $\eta(\pi_a) < 1.5$;                                            
\item  $p_T (D^{+}\pi_a)/E_{\perp}^{\theta > 10^{\circ}} > 0.35$;  
\item  $\cos\theta^*(D^{+}) < 0.8$, where
$\theta^*(D^{+})$ is the angle between the $D^{+}$ in the $D^{+}\pi_a$ rest frame and the $D^{+}\pi_a$ line of
flight in the laboratory frame;             
\item  the cut $l_{\pi} > 0.01$ was applied for $\pi_a$.   
 \end{itemize}
The $D^*_2(2460)^0\to D^{+}\pi^-$  decay mode was reconstructed by calculating 
                                        the ``extended'' mass difference  
$\Delta M^{\rm{ext}} = M(K\pi\pi\pi_a) - M(K\pi\pi)$. 
                            Figure~\ref{3}~(b) shows the invariant mass 
$M(D^{+}\pi_a) = \Delta M^{\rm{ext}} + M(D^{+}_{\mathrm{PDG}})$, where $M(D^{+}_{\mathrm{PDG}})$ is the nominal $D^{+}$
mass~\cite{PDG11}.                                                                                                           
        A clear $D^{*0}_2$ signal is seen.                                   
No indication of the $D^0_1\to D^+\pi^-$ decay is seen, as expected from angular momentum and parity conservation
for a $J^P = 1^+$ state. 
The various contributions to the mass spectrum will be discussed below.
 
\subsection{Reconstruction of the $\boldsymbol{D_1 (2420)^{+}}$ and $\boldsymbol{D^{*}_2 (2460)^{+}}$ mesons}
 \label{sec-dsstwo}
The charged excited meson $D_1 (2420)^{+}$ has been seen~\cite{PDG11} in the decay modes  $D^{*0}\pi^{+}$ and  $D^{+}\pi^+\pi^-$ and
the charged excited meson $D^{*}_2 (2460)^{+}$ has been seen~\cite{PDG11} in the decay modes  $D^{*0}\pi^{+}$ and  $D^0\pi^{+}$.
A search for $D^+_1$ and $D^{*+}_2$ signals was performed in the mass distribution $M(D^0\pi^{+})$.
For the $D^+_1$ a possible $D^0\pi^+$ signal can arise only via a feed-down
contribution (see Section~\ref{sec-mwhel}).
        Each $D^{0}$ candidate 
                                                            was combined with an additional track, assumed to be
a pion ($\pi_a$), with either positive or negative charge.
Combinatorial background was reduced by applying the following cuts:   
 \begin{itemize}
\item  $p_T (\pi_a) > 0.35$~GeV; 
\item  $\eta(\pi_a) < 1.6$;                                             
\item  $p_T (D^{0}\pi_a)/E_{\perp}^{\theta > 10^{\circ}} > 0.3$;                              
\item  $\cos\theta^*(D^{0}) < 0.85$, where
$\theta^*(D^{0})$ is the angle between the $D^{0}$ in the $D^{0}\pi_a$ rest frame and the $D^{0}\pi_a$ line of
flight in the laboratory frame;   
\item  the cut $l_{\pi} > 0.01$ was applied for $\pi_a$.   
 \end{itemize}
For each excited charm-meson candidate,                                         
                                        the ``extended'' mass difference 
$\Delta M^{\rm{ext}}\,=\,M(K\pi\pi_a)-M(K\pi)$ was calculated.
                            Figure~\ref{4} shows the invariant mass 
$M(D^{0}\pi_a) = \Delta M^{\rm{ext}} + M(D^{0}_{\mathrm{PDG}})$, where $M(D^{0}_{\mathrm{PDG}})$ is the nominal $D^{0}$
mass~\cite{PDG11}.
A clear signal of $D^{*+}_2\to D^0\pi^{+}$ is seen. 
An enhancement above background is also seen at the mass region around $2.3$~GeV.
The various contributions to the mass spectrum will be discussed below.
 
\section{Mass, width and helicity parameters of $\boldsymbol{D_1}$ and $\boldsymbol{D^*_2}$ }
 \label{sec-mwhel}
 
A significant enhancement above background is seen  
in the                     $D^{0}\pi^{+}$ mass distribution  (Fig.~\ref{4}) around $2.3$~GeV.
                                                          A small excess of events is also seen in the same
mass region in the $D^{+}\pi^{-}$ mass distribution (Fig.~\ref{3}(b)). 
 
The origin of these structures in both spectra is      similar. They originate from 
                   the decay chains   
$D^{0}_{1},D^{*0}_{2} \rightarrow D^{*+}\pi^{-},\ D^{*+}  \rightarrow D^{+} \pi^{0} $ and
$D^{+}_{1},D^{*+}_{2} \rightarrow D^{*0}\pi^{+}, D^{*0}  \rightarrow D^{0} \pi^{0} $ or
$D^{*0}  \rightarrow D^{0}\gamma$.
The $\pi^{0}/\gamma$ are not seen in the tracking detectors; thus, the            
                                                  reconstruction is incomplete.
However, since the available phase space in the
  $D^{*} \rightarrow D \pi^{0} $ decay  is small and $D$ is much heavier than $\pi^{0}$,
the energy and momentum of $D$ are close to those of $D^{*}$.  
Consequently, the enhancements in the $M(D^{+}\pi_a)$ (Fig.~\ref{3}(b)) and
                                      $M(D^{0}\pi_a)$ (Fig.~\ref{4}) distributions are feed-downs of the
excited charm mesons $D_1 , D^*_2$, shifted down approximately by the  value of the $\pi^0$ mass,
as verified by MC simulations.

\subsection{Fitting procedure for $\boldsymbol{D^0_1}$ and $\boldsymbol{D^{*0}_2}$ }
 
To distinguish between $D^0_1,D^{*0}_2\to D^{*+}\pi^{-}$,
                 their                        helicity angular distributions were used.
These can be parametrised as $dN/d\cos\alpha~ \propto 1+h\cos^2\alpha$, where
 $\alpha$  is the angle between the $\pi_a$ and $\pi_s$ momenta
in the $D^{*+}$ rest frame and $h$ is the helicity parameter, predicted~\cite{HQET1,HQET2} to be $h=3$
for $D^0_1$ and $h = -1$ for $D^{*0}_2$. 
Figure~\ref{5} shows the $M(D^{*+}\pi_a)$ distribution in four
helicity bins.
As expected from the above $h$ values, the $\done$ 
contribution increases with $|\cos\alpha|$ and dominates for $|\cos\alpha| > 0.75$,
where the $\dtwo$ contribution is negligible.
 
A  $\chi^2$ fit   
was performed using  simultaneously                 
   the $M(D^{+}\pi_a)$ distribution (Fig.~\ref{3}(b)) and the 
$M(D^{*+}\pi_a)$ distributions in four helicity bins (Fig.~\ref{5}).
The background was described by four parameters $a,b,c,d$, 
separately for $M(D^{*+}\pi_a)$ and $M(D^{+}\pi_a)$, as
$B(x) = ax^b\exp(-cx-dx^2)$, where $x = \Delta M^{\rm{ext}} - M_{\pi^+}$. 
Each resonance was fitted to a relativistic $D$-wave Breit-Wigner (BW) function~\cite{dsshera1}         convoluted with a Gaussian
resolution function with a width fixed to the corresponding MC prediction.                   
Yields of the three signals, the $D^0_1$ and $D^{*0}_2$ masses and widths
                              and the $D^0_1$ helicity parameter, $h(D^0_1)$,
 were free parameters of the fit while
    $h(D^{*0}_2)$ was fixed to the theoretical
 prediction~\cite{HQET1,HQET2}, $h(D^{*0}_2) = -1$.                                                         
Another free fit parameter was the contribution of the $D^{0}_{1},D^{*0}_{2}$ feed-downs to the $M(D^{+}\pi_a)$ distribution (see Appendix).
The total normalisation of the sum of the feed-down processes from $D^{*0}_2$ and  $D^0_1$ decays
was fitted relative to the direct signal peak yield from $D^{*0}_2$ decay.
The relative yields of the two feed-down contributions were 
taken to be equal to those for the direct signals in the $D^{*+}\pi^-$ decay channel.

The wide excited charm  states~\cite{PDG11} $D_1(2430)^0$ 
and $D^*_0(2400)^0$ are expected to contribute 
              to the $M(D^{*+}\pi_a)$ and $M(D^{+}\pi_a)$ distributions,
  respectively. Even though these states are hardly distinguishable
from background due to their large width, they were included in the simultaneous fit with shapes
                                                                            described as relativistic
$S$-wave BW functions~\cite{dsshera1}. Their masses and widths were set to the PDG values~\cite{PDG11}. The yield of the 
$D_1(2430)^0$ was set to that of the narrow $D_1(2420)^0$ meson since both have the same spin-parity $J^P = 1^+$.
The ratio of $D^*_0(2400)^0$ to the narrow state $D^{*}_2(2460)^0$ was a free parameter in the fit.                                               
 
 The results of the simultaneous fit                                       
 are given in Table~\ref{tab2} and shown in Figs.~\ref{3} and ~\ref{5}.
 Systematic uncertainties are discussed in Section~\ref{sec-sys}.
 All results from the new analysis (HERA II) are consistent with those
from the previous ZEUS publication~\cite{dsshera1} (HERA I).
 The masses of both $D^0_1$ and $D^{*0}_2$ are consistent with the PDG values~\cite{PDG11} and with     
 a recent BABAR measurement~\cite{Babar}.
The $D^0_1$ width, $\Gamma(D^0_1) =~$\DZeroOneWidthStatSyst~MeV, is also            
   consistent with the PDG value~\cite{PDG11} of $27.1 \pm 2.7$~MeV, and is in good agreement                    
                                      with the BABAR measurement~\cite{Babar}
of $31.4 \pm 0.5 \pm 1.3$~MeV.                                                        
The $D^{*0}_2$ width, $\Gamma(D^{*0}_2) =~$\DZeroStarTwoWidthStatSyst~MeV,                    
is  consistent with the PDG value~\cite{PDG11} of $49.0 \pm 1.4$~MeV, and                     
                                                                               with the BABAR measurement              
of $50.5 \pm 0.6 \pm 0.7$~MeV.                                                        
 
The $D^0_1$ helicity parameter,                                       
$h(D^0_1) =\,$\DZeroOneHelicityStatSyst
, is consistent with the BABAR value of 
$h(D^0_1) = 5.72 \pm 0.25$ and somewhat above the theoretical 
prediction of $h = 3$ and measurements by CLEO~\cite{CLEO} with $h(D^0_1) = 2.74^{+1.40}_{-0.93}$.
The simultaneous fit with $h(D^0_1)$ fixed to the theoretical prediction, $h(D^0_1) = 3$,        
     yielded masses and widths of $D^{*0}_2$ and $D^{0}_1$ that are somewhat away from the PDG values~\cite{PDG11}.
Repeating the simultaneous fit with $h(D^{*0}_2)$ as a free parameter yielded similar results for all other free
parameters with somewhat larger errors and with $h(D^{*0}_2) = $\DZeroStarTwoHelicity, in good agreement
with the theoretical prediction of $h = -1$.

The helicity angular distribution for a $J^P = 1^+$ state with a mixture of $D$- and $S$-wave is
\begin{equation}
\frac {dN}{d\cos\alpha} \propto r + (1-r)(1+3\cos^2\alpha)/2 + \sqrt{ 2r(1-r)} \cos\phi(1-3\cos^2\alpha),
\end{equation}
where  $ r=\Gamma_S/(\Gamma_S + \Gamma_D)$,               $\Gamma_{S}(\Gamma_{D})$
  is the $S(D)$-wave partial width and
$\phi$  is relative phase between the two amplitudes. The relation between $h$, $r$ and $\phi$
is given by 
\begin{equation}
\cos\phi = \frac {(3-h)/(3+h) - r} {2\sqrt{2r(1-r)}}.
\end{equation} 
The range of the measured $h(D^0_1)$ restricted to
    one standard deviation                                              is 
shown in Fig.~\ref{6} in a plot of
$\cos\phi$ versus $r$. 
 This range is consistent with the BABAR measurement\cite{Babar}. 
The range restricted by CLEO~\cite{CLEO}     
 is outside the range of this measurement and that of BABAR.                                       
                                                                    A similar measurement by the                                     
  BELLE collaboration~\cite{Belle} is consistent with a pure $D$-wave, i.e. $\Gamma_S/(\Gamma_S + \Gamma_D)=0$.
                                                                                              
 
In a recent paper~\cite{Babar} the BABAR Collaboration searched for excited $D$ meson states in
$e^+ e^- \to c\bar c\to D^{(*)}\pi + X$ with very large statistics. In addition to the $D^0_1$ and
$D^{*0}_2$ resonances, they saw two new structures near $2.6$~GeV in the $D^+\pi^-$ and 
$D^{*+}\pi^-$ mass distributions, $D(2550)^0$ and $D^*(2600)^0$, and interpreted them as being radial
excitations of the well-known $D^0$ and $D^{*0}$, respectively. 
 A small enhancement of events above the solid curve in the region
near $~2.6$~GeV is seen
            in the $M(D^{*+}\pi^-)$ distribution                                     
              (Figs.~\ref{3}(a),\ref{5}).                                                       
Adding the new BABAR states to the fit gave insignificant
yields of the states and did not significantly change the results of the
other fit parameters.

\subsection{Fitting procedure for $\boldsymbol{D^+_1}$ and $\boldsymbol{D^{*+}_2}$ }
 \label{sec-fitcharge}
To extract the $D^+_1$ and $D^{*+}_2$ masses         and yields, a minimal $\chi^2$ fit   
was performed using                                 
   the $M(D^{0}\pi_a)$ distribution (Fig.~\ref{4}).        
Both resonances were fitted to   relativistic $D$-wave Breit-Wigner (BW) functions~\cite{dsshera1}         convoluted with a Gaussian
resolution function with a width fixed to the corresponding MC prediction.                   
Yields of the $D^{*+}_2\to D^0\pi^{+}$ and the two feed-downs $D^+_1 , D^{*+}_2\to D^{*0}\pi^{+}$ 
(see Appendix) and
                               the $D^+_1$ and $D^{*+}_2$ masses               
 were free parameters of the fit.                                          
                  The $D^+_1$ and $D^{*+}_2$ widths were fixed to the PDG values~\cite{PDG11} and
 the $D^+_1$ and $D^{*+}_2$ helicities  were fixed to the theoretical
 prediction~\cite{HQET1,HQET2}, $h(D^{+}_1) = 3$ and $h(D^{*+}_2) = -1$.                                                         
The background was parametrised with four parameters $a,b,c,d$ as 
$B(x) = a x^b \exp(-cx-dx^2)$, where $x = \Delta M^{\rm{ext}} - M_{\pi^+}$.

    The results of the              fit (yields and masses)                   
 are given in Table~\ref{tab3} and shown in Fig.~\ref{4}.
 The masses of $D^+_1$ and $D^{*+}_2$ are consistent with the  PDG values~\cite{PDG11}. The $D^{*+}_2$ mass is 
also consistent with the BABAR measurement~\cite{Babar}.

\section{$\boldsymbol{D_2^*}$ decay branching ratios and $\boldsymbol{D_1/D_2^*}$ fragmentation fractions}
 
\subsection{ 
The neutral excited mesons}
\label{frag-neutral} 
The branching ratio     for  $\dtwo$ and the fragmentation fractions for $\done$ and $\dtwo$
were measured using the channels $\dtwo\rightarrow D^{+}\pi^-$  and  $\done,\dtwo\rightarrow D^{*+}\pi^-$
         with $D^{*+}\to D^0\pi^+_s\to (K^-\pi^+)\pi^+_s$.
The numbers of reconstructed $\done,\dtwo\rightarrow D^{*+}\pi^-$ and
$\dtwo\rightarrow D^{+}\pi^-$ decays were
divided by the numbers of reconstructed $\dsp$ and $\dc$ mesons,
yielding the fractions of $\dsp$ and $\dc$ mesons originating from
the $\done$ and $\dtwo$ decays.
To correct the measured fractions for detector effects,
ratios of acceptances were calculated using the MC simulation
for the  $\done,\dtwo\rightarrow D^{*+}\pi^-$ and
$\dtwo\rightarrow D^{+}\pi^-$ states to the inclusive $\dsp$
and $\dc$ acceptances, respectively.
 
Beauty production at HERA is smaller than charm production
by two orders of magnitude.
A subtraction of 
 the $b$-quark relative contribution
 in a previous ZEUS analysis~\cite{dsshera1} 
                                  changed the relative acceptances
 by less than 1.5\% of their values.
Consequently, no such subtraction was performed in this analysis and
 the MC simulation                       included
      the beauty production processes.
  A variation of  this contribution was considered
for the systematics (Section~\ref{sec-sys}).
 
The fractions, $\fr$, of $\dsp$ mesons
originating from $\done$ and $\dtwo$ decays were
calculated in the kinematic range
$|\eta(D^{*+})|<1.6$
and
$p_T(D^{*+})>1.5\,$GeV
for the $\dsp$ decay
and the fraction of $\dc$ mesons
originating from $\dtwo$ decays was
calculated in the kinematic range
$p_T(D^{+})>2.8\,$GeV and $|\eta(D^{+})|<1.6$.
 
The fractions measured in the restricted
$p_T(D^{*+},D^+)$ and $\eta(D^{*+},D^+)$
kinematic ranges were extrapolated to the fractions in the full kinematic
phase space
using the Bowler modification~\cite{zfp:c11:169}
of the Lund symmetric fragmentation function~\cite{zfp:c20:317}
as implemented in
{\sc Pythia}~\cite{cpc:82:74}.
Applying the estimated extrapolation factors,
$\sim 1.12$ for $\fr_{D_1^0\rightarrow D^{*+}\pi^-/D^{*+}}$,
$\sim 1.16$ for $\fr_{D_2^{*0}\rightarrow D^{*+}\pi^-/D^{*+}}$
and $\sim 1.34$ for $\fr_{D_2^{*0}\rightarrow D^{+}\pi^-/D^{+}}$, gives

\begin{equation}
\fr^{\rm extr}_{D_1^0\rightarrow D^{*+}\pi^-/D^{*+}}
=\DZeroOneExtrapolationDStarStatSyst,
\end{equation}

\begin{equation}
\fr^{\rm extr}_{D_2^{*0}\rightarrow D^{*+}\pi^-/D^{*+}}
=\DZeroStarTwoExtrapolationDStarStatSyst,
\label{fdstar}
\end{equation}

\begin{equation}
\fr^{\rm extr}_{D_2^{*0}\rightarrow D^{+}\pi^-/D^{+}}
=\DZeroStarTwoExtrapolationDPlusStatSyst.
\label{fdplus}
\end{equation}
 
 In the full kinematic phase space,
 the extrapolated fractions of
 $\dsp$ originating from $\done$ and $\dtwo$
 and of $\dc$ originating from $\dtwo$
 can be expressed~\cite{dsshera1} in terms of the rates of $c$-quarks hadronising to a given charm meson
(``fragmentation fractions''),
                                   $\fcdone$, $\fcdtwo$, $\fcds$ and $\fcdc$
and the corresponding branching fractions\linebreak
     $\bran_{D^0_1\rightarrow D^{*+}\pi^-}$,
 $\bran_{D^{*0}_2\rightarrow D^{*+}\pi^-}$ and
 $\bran_{D^{*0}_2\rightarrow D^{+}\pi^-}$.                                  
  
 From the expressions used in a previous ZEUS publication~\cite{dsshera1}, 
the fragmentation fractions $\fcdone$ and $\fcdtwo$  and
the ratio of the
two branching fractions
for the $\dtwo$ meson can be shown to be:
\begin{equation}
\label{eq:fcd1}
\fcdone =
\frac{\fr^{\rm extr}_{D_1^0\rightarrow D^{*+}\pi^-/D^{*+}}}
{\bran_{D^0_1\rightarrow D^{*+}\pi^-}}\fcds,
\end{equation}
\begin{equation}
\label{eq:fcd2}
\fcdtwo = 
\frac{\fr^{\rm extr}_{D_2^{*0}\rightarrow D^{*+}\pi^-/D^{*+}}\fcds
     +\fr^{\rm extr}_{D_2^{*0}\rightarrow D^{+}\pi^-/D^{+}}\fcdc}
{\bran_{D_2^{*0}\rightarrow D^{*+}\pi^-}+\bran_{D_2^{*0}\rightarrow D^{+}\pi^-}},
\end{equation}
\begin{equation}
\frac{\bran_{D_2^{*0} \rightarrow D^+ \pi^-}}{\bran_{D_2^{*0} \rightarrow D^{*+} \pi^-}} =
\frac{\fr^{\rm extr}_{D_2^{*0}\rightarrow D^{+}\pi^-/D^{+}}\fcdc}
{\fr^{\rm extr}_{D_2^{*0}\rightarrow D^{*+}\pi^-/D^{*+}}\fcds}\,.
\end{equation} 
The $\fcds$ and $\fcdc$ values used were obtained as a combination of data from HERA and $e^{+}e^{-}$ colliders\cite{lohrmann}:
$$\fcds = 22.87 \pm 0.56(\rm stat.\oplus \rm syst.)^{+0.45}_{-0.56}(\rm br.)\,\%,$$
$$\fcdc = 22.56 \pm 0.77(\rm stat.\oplus \rm syst.)\pm 1.00 (\rm br.)\,\%.$$
where the third uncertainties are due to the branching-ratio uncertainties.
 
Taking into account the correlations in the simultaneous fit  performed to obtain 
the values in     
Eqs.~(\ref{fdstar}) and (\ref{fdplus})
 yields
$$\frac{\bran_{D_2^{*0} \rightarrow D^+ \pi^-}}
{\bran_{D_2^{*0} \rightarrow D^{*+} \pi^-}} =
\DZeroStarTwoBranchingStatSyst,$$
in good      agreement with the PDG world-average value
   $1.56 \pm 0.16$~\cite{PDG11}.
Theoretical models~\cite{cnpp:16:109,pr:d43:1679,pr:d49:3320}
predict the ratio to be in the range
from $1.5$ to $3$.
 
Neglecting the contributions of the      
   non-dominant decay mode $D^0_1\to D^0\pi^+\pi^-$~\cite{PDG11} and assuming isospin conservation, for which
$$\bran_{D_1^{0} \rightarrow D^{\ast +} \pi^-} = 2/3,
\,\,\,\,\bran_{D^{*0}_2\rightarrow D^{*+}\pi^-}+
\bran_{D^{*0}_2\rightarrow D^{+}\pi^-} = 2/3,$$
and using Eqs.~(\ref{eq:fcd1}) and (\ref{eq:fcd2})
yields 
$$\fcdone = \DZeroOneFragmentationStatSyst,$$
$$\fcdtwo = \DZeroStarTwoFragmentationStatSyst.$$
The measured fragmentation fractions were
             found to be
consistent with
those obtained in $e^+e^-$ annihilations~\cite{epemone}.
The sum of the two fragmentation fractions,
$$\fcdone + \fcdtwo = \DExcitedNeutralFragmentationSumStatSyst,$$
agrees with the prediction
of the tunneling model
of $8.5\%$~\cite{zfp:c72:39}.
 
Assuming uncorrelated errors,
the ratio                                                       
$$\fcdone/\fcdtwo = \DExcitedNeutralFragmentationRatioStatSyst$$
is in good      agreement with the simple spin-counting prediction of $3/5$.
 
\subsection{ 
The charged excited mesons}
 
The  branching ratio for $D^{*+}_{2}$ and the fragmentation fractions for $D^+_1$ and $D^{*+}_2$
were measured using the channels $D^{*+}_{2}\rightarrow D^{0} \pi^{+}$  and 
$D^+_1 , D^{*+}_{2}\rightarrow D^{*0} \pi^{+}$ with $D^{*0} \rightarrow D^{0} \pi^{0}/\gamma$, 
where the $\pi^{0}/\gamma$ are not measured directly. Since
      $D^{*0}$ decays always  to $D^0$\cite{PDG11}, the number of $D^{*0}$ and $D^0$
originating from $D^+_1 / D^{*+}_2$ are identical.
                                                                The number of reconstructed 
$D^+_1 / D^{*+}_2\to D^{*0} \pi^{+}; D^{*0}\to D^0\pi^0 /\gamma$ and $D^{*+}_2\to D^{0}\pi^{+}$ decays
were thus divided by the total number of reconstructed $D^0$ mesons, yielding the fractions of $D^0$ mesons 
originating from $D^+_1 / D^{*+}_2$ decays. Detector effects were corrected as described in Section~\ref{frag-neutral}.
The above fractions were calculated in the kinematic range
                    $p_T(D^{0})>2.6$~GeV and $|\eta(D^{0})|<1.6$ 
and extrapolated to the fractions in the full kinematic
phase space     
           as for the $D^0_1$ and $D^{*0}_2$ (Section~\ref{frag-neutral}).
Applying the extrapolation factors, $\sim 1.28$ for $D^{+}_{1}\rightarrow D^{*0} \pi^{+}$, 
$\sim 1.18$ for $D^{*+}_{2}\rightarrow D^{*0} \pi^{+}$ and $\sim 1.35$ for $D^{*+}_{2}\rightarrow D^{0} \pi^{+}$ gives

\begin{equation}
   \fr^{\rm extr}_{D_1^+\rightarrow D^{*0}\pi^+/D^{0}}=\DPlusOneExtrapolationDZeroStarStatSyst,  
\label{DonetoDstarZ}
\end{equation}
 
\begin{equation}
   \fr^{\rm extr}_{D_2^{*+}\rightarrow D^{*0}\pi^+/D^{0}}=\DPlusStarTwoExtrapolationDZeroStarStatSyst, 
\label{DtwotoDstarZ}
\end{equation}
 
\begin{equation}
  \fr^{\rm extr}_{D_2^{*+}\rightarrow D^{0}\pi^+/D^{0}}=\DPlusStarTwoExtrapolationDZeroStatSyst.  
\label{DtwotoDZ}
\end{equation}

The fractions of $D^{*0}/D^0$ mesons  originating from $D^+_1 / D^{*+}_{2}$ decays can be expressed as
 
\begin{equation}
  \fr^{\rm extr}_{D_1^+\rightarrow D^{*0}\pi^+/D^{*0}}\equiv\frac {N(D_1^+\rightarrow D^{*0}\pi^+)}{N(D^{*0}) } = 
\frac {\fcdoneplus} {f(c\to D^{*0})} \bran_{D_1^{+} \rightarrow D^{*0} \pi^+},  
\label{rate1}
\end{equation}
 
\begin{equation}
  \fr^{\rm extr}_{D_2^{*+}\rightarrow D^{*0}\pi^+/D^{*0}}\equiv\frac {N(D_2^{*+}\rightarrow D^{*0}\pi^+)}{N(D^{*0}) } = 
\frac {\fcdtwoplus} {f(c\to D^{*0})} \bran_{D_2^{*+} \rightarrow D^{*0} \pi^+},  
\label{rate2}
\end{equation}
 
\begin{equation}
  \fr^{\rm extr}_{D_2^{*+}\rightarrow D^{0}\pi^+/D^{0}}\equiv\frac {N(D_2^{*+}\rightarrow D^{0}\pi^+)}{N(D^{0}) }=
\frac {\fcdtwoplus} {f(c\to D^{0})} \bran_{D_2^{*+} \rightarrow D^{0} \pi^+ },  
\label{rate3}
\end{equation}
 where $N$ denotes the acceptance-corrected number of events.
 
The ratio of the fragmentation fractions $f(c\to D^{*0})$ and $f(c\to D^{0})$ can be expressed as
 
$$\frac {f(c\to D^{*0})} {f(c\to D^{0})} =                            
  \frac {N(D^{*0})} {N(D^{0})}.  
$$
 
                                Consequently, Eqs.~(\ref{rate1}) and (\ref{rate2}) can be written as
 
$$\frac {N(D_1^+\rightarrow D^{*0}\pi^+)} {N(D^0) } = 
\frac {\fcdoneplus} {f(c\to D^{0})} \bran_{D_1^{+} \rightarrow D^{*0} \pi^+}, 
$$
 
$$\frac  {N(D_2^{*+}\rightarrow D^{*0}\pi^+)}{N(D^0) } = 
\frac {\fcdtwoplus} {f(c\to D^{0})} \bran_{D_2^{*+} \rightarrow D^{*0} \pi^+}, 
$$
 
yielding 
 
 $$f(c\to D^{+}_1)=\frac {f(c\to D^{0})} {N(D^0)}         
                   \frac  {N(D_1^+\rightarrow D^{*0}\pi^+)} {\bran_{D_1^{+} \rightarrow D^{*0} \pi^+} },
 $$

 $$f(c\to D^{*+}_2)=\frac {f(c\to D^{0})} { N(D^0) }
                   \frac {N(D_2^{*+}\rightarrow D^{*0}\pi^+) + N(D_2^{*+}\rightarrow D^{0}\pi^+)}                         
                         {\bran_{D_2^{*+} \rightarrow D^{*0} \pi^+} + \bran_{D_2^{*+} \rightarrow D^{0} \pi^+} },  
 $$

 $$\frac {\bran_{D_2^{*+} \rightarrow D^{0} \pi^+}}        {\bran_{D_2^{*+} \rightarrow D^{*0} \pi^+}} =
   \frac {N(D_2^{*+}\rightarrow D^{0}\pi^+)}{N(D_2^{*+}\rightarrow D^{*0}\pi^+)}~.
 $$

Neglecting the non-dominant decay mode $D^+_1\to D^+\pi^+\pi^-$~\cite{PDG11}, assuming isospin conservation, for which
$$\bran_{D_1^{+} \rightarrow D^{\ast 0} \pi^+} = 2/3,
\,\,\,\,\bran_{D^{*+}_2\rightarrow D^{*0}\pi^+}+
\bran_{D^{*+}_2\rightarrow D^{0}\pi^+} = 2/3,$$
and using Eqs.~(\ref{DonetoDstarZ} -- \ref{DtwotoDZ})                                
and the fragmentation fraction\cite{lohrmann} 
$$\fcdz = 56.43 \pm 1.51(\rm stat.\oplus \rm syst.)^{+1.35}_{-1.64}(\rm br.)\,\%,$$
gives          
 
$$\fcdoneplus = \DPlusOneFragmentationStatSyst,$$
 
$$\fcdtwoplus = \DPlusStarTwoFragmentationStatSyst,$$
 
$$\fcdoneplus+\fcdtwoplus = \DExcitedChargedFragmentationSumStatSyst,$$
 
$$\fcdoneplus/\fcdtwoplus = \DExcitedChargedFragmentationRatioStatSyst,$$

in agreement with the fragmentation fractions of the neutral excited charm mesons (Section~\ref{frag-neutral}). 
 
The ratio of the branching fractions of the two dominant decay modes of the $D^{*+}_2$, 
 
\begin{equation}
\frac{\bran_{D_2^{*+} \rightarrow D^0 \pi^+}}{\bran_{D_2^{*+} \rightarrow D^{*0} \pi^+}} = \DPlusStarTwoBranchingStatSyst, 
\label{dtworatio}
\end{equation}
significantly improves on the accuracy of the PDG~\cite{PDG11} value of $1.9 \pm 1.1 \pm 0.3$.
                               BABAR measured the ratio~\cite{prl:103:051803} 
$\frac {\bran_{D_2^{*+} \rightarrow D^{0} \pi^+}} {\bran_{D_2^{*+} \rightarrow D^{0} \pi^+} + \bran_{D_2^{*+} \rightarrow D^{\ast 0} \pi^+}} = 0.62 \pm 0.03 \pm 0.02$,
 which depends on some assumptions and is        
not included in the PDG averages~\cite{PDG11}.
Using the value given in Eq.(\ref{dtworatio}) yields a ratio
$\frac {\bran_{D_2^{*+} \rightarrow D^{0} \pi^+}} {\bran_{D_2^{*+} \rightarrow D^{0} \pi^+} + \bran_{D_2^{*+} \rightarrow D^{\ast 0} \pi^+}} = 
\DPlusStarTwoBranchingBABARStyleStatSyst$, in good agreement with the BABAR result.

\section{Systematic uncertainties              }
\label{sec-sys}
 
 The systematic uncertainties were
determined by appropriate variations of the analysis procedure, generally by the uncertainties in our knowledge of the variables
considered,
and repeating the calculation of the results.
The following sources of uncertainty were considered:

\begin{itemize}
\item \{$\delta_1$\} 
 The stability of the fit results was checked
 by a variation of the selection cuts which are most sensitive
 to the ratio of signal and background in the data:
    \begin{itemize}
\item the cut on the minimal transverse momentum of the 
 $D^{*+}$, $D^{+}$ and $D^0$ candidates was varied by $\pm 100\,$MeV;
\item the cut on the  minimal transverse momentum of the extra pion in the excited $D$ meson
analysis was varied by $\pm 10\,$MeV;
\item[--]  the selection cut on the cosine of angle between extra pions and charged (neutral) excited $D$ meson candidates was changed by $\pm 0.1$  
              ($\pm 0.05$);
\item[--] the widths of the mass windows used for the selection of $D^{*+}$
and $D^{0}$ candidates in the excited charm meson analyses were varied by $\pm 5\%$ for each $p_T$ dependent window
(see Table~\ref{tab1}), 
while for the $D^{+}$ candidates it was varied by $\pm 12.5\%$.
    \end{itemize}
 
\item \{$\delta_2$\} 
The CAL energy scale is known with $\pm2\%$ uncertainty
and was varied accordingly in the simulation.
 
\item \{$\delta_3$\} The uncertainties related to the fit  procedure
were obtained as follows:

    \begin{itemize}
    \item[--] the ranges for the signal fits
were reduced on either side  by 16~MeV for the $D^{*+}\pi$ and $D^+\pi$ mass spectra and 24~MeV for
the $D^0\pi$ mass spectrum;
    \item[--]  the background shape was changed to that used by BABAR (Eq.~1 in ref.~\cite{Babar}); 
    \item[--] the widths of the Gaussians used to parametrise the mass resolutions
were changed by $\pm20\%$;
    \item [--] 
all the masses and widths of wide states were set free in the fit.
Since with the present data alone these parameters are not determined well,
the world-average values from  PDG~\cite{PDG11} were used as additional constraints.
This was implemented by adding for each parameter $P$ (width or mass) a 
term ${\frac{(P-P_{\mathrm{PDG}})^2}{\sigma (P_{\mathrm{PDG}})^2}}$ to the $\chi^2$-function.
Here $P_{\mathrm{PDG}}$ and $\sigma (P_{\mathrm{PDG}})$ denote the parameter value and its uncertainty from PDG~\cite{PDG11};
    \item [--] the background functions in the four helicity
intervals were allowed to have separate normalisations;

    \item[--] the helicity parameter of the $\mbox{$D_2^{\ast 0}$}$ meson in the fit 
was  set free (Section~\ref{sec-dssone}). 
 
    \end{itemize}
\item \{$\delta_4$\} The uncertainties of $M(\dsp)_{\mathrm{PDG}}$, $M(\dz)_{\mathrm{PDG}}$, $M(\dc)_{\mathrm{PDG}}$ were taken into account.
\item \{$\delta_5$\} The widths of $D^{+}_{1}$ and $D^{*+}_{2}$ 
 were varied within their uncertainties taken from PDG~\cite{PDG11}.
 
\item \{$\delta_6$\} 
The uncertainty of the beauty contamination was determined
by varying the    beauty fraction in the MC sample between $0$ and $200\%$ of the reference amount.
\item \{$\delta_7$\} 
The extrapolation uncertainties were determined 
by varying relevant parameters of the {\sc Pythia} simulation
using the Bowler modification~\cite{zfp:c11:169}
of the Lund symmetric fragmentation function~\cite{zfp:c20:317}.
The following variations were performed:
\begin{itemize}
\item{
the mass of the $c$ quark was varied from its nominal value of $1.5$~GeV by $\pm 0.2$~GeV;
}
\item{
the strangeness suppression factor was varied from its nominal value of $0.3$ by $\pm 0.1$;
}
\item{
the fraction of the lowest-mass charm mesons produced in a vector state
was varied from its nominal value of $0.6$ by $\pm 0.1$;
}
\item{
the Bowler fragmentation function parameter $r_c$ was varied from
the predicted value $1$ to $0.5$;
the $a$ and $b$ parameters of the Lund symmetric function
were varied by $\pm20\%$
around their default values~\cite{cpc:82:74}.
}
\end{itemize}
\end{itemize}
 A possible model dependence of the acceptance corrections was checked
by reweighting the D-meson transverse momentum distribution 
in the MC to match the distribution observed in the data; no significant effect
on any result was found. 
As a further cross check the selected pseudorapidity range of the extra pion, 
which is not the same
for the different decay channels (see Section 5), was varied,
and again no significant effect on any result was observed.
The uncertainties of the fragmentation fractions
$f(c\rightarrow \dsp)$, $f(c\rightarrow \dc)$ and $f(c\rightarrow \dz)$ 
were included by adding in quadrature their statistical and systematic
uncertainties and the uncertainties originating from the
branching-ratio  uncertainties. The resulting uncertainty 
is included in $\delta_7$.
 
The contributions from all systematic uncertainties were calculated separately 
for positive and negative        variations and added in quadrature.
The obtained values are listed in 
Tables~\ref{tab:syst_mg}--\ref{tab:syst_charged}.
There is no single dominating source of systematic uncertainty.
The total systematic uncertainties are    comparable 
     to the statistical errors.
\section{Summary}                      
\label{sec-con}
 
The full HERA data taken from 2003 to 2007 with 
an integrated luminosity of $373$ pb$^{-1}$ has been used to study the production of 
 excited charm
mesons.
~Signals of $D_1(2420)^0$ and $D^*_2(2460)^0$
were  seen in the  $D^{*+}\pi^{-}$ decay mode and a clear $D^*_2(2460)^0$ signal was seen in the
$D^{+}\pi^{-}$ decay mode. The measured $D_1^0$ and $D^{*0}_2$ masses and widths are in
good agreement with the latest PDG values. 
The
measured $D_1^0$ helicity parameter allows for some $S$-wave mixing in its decay
                                 to $D^{*+}\pi^{-}$.
    The result is also consistent with a pure $D$-wave hypothesis. The helicity of
  $D^{*0}_{2}$, when set free in the fit,
 is consistent with the HQET prediction, $h = -1$. 
 
A clear $D^*_2(2460)^{+}$ signal is seen for the first time at HERA in the 
$D^{0}\pi^{+}$ decay mode. Feed-downs of both resonances $D_1(2420)^{+}$ and $D^*_2(2460)^{+}$
in the decay mode $D^{*0}\pi^{+}$ are seen in the expected mass region of $M(D^{0}\pi^{+})\approx 2.3$~GeV.
The measured $D^+_1$ and $D^{*+}_2$ masses are in good agreement with the PDG                 
 values and the
$D^{*+}_2$ mass is consistent with the BABAR measurement.
 
The fractions of $c$-quarks hadronising into $D^0_1$ and $D^{*0}_2$ are consistent with those from the previous ZEUS
publication                   
and with $e^+ e^-$ annihilation results, in agreement with charm fragmentation universality.
The fractions of $c$-quarks hadronising into $D^+_1$ and $D^{*+}_2$ were measured for the first time         and are
consistent, respectively,  with the fractions of the neutral charm excited states $D^0_1$ and $D^{*0}_2$.
 
The ratios of the neutral and charged $D^*_2$ branching ratios into $D\pi$ and $D^*\pi$ are consistent with
the  PDG values.
 
\section*{Acknowledgements}
\label{sec-ack}
 
We appreciate the contributions to the construction and maintenance of
the ZEUS detector of many people who are not listed as authors. The
HERA machine group and the DESY computing staff are especially
acknowledged for their success in providing excellent operation of the
collider and the data-analysis environment. We thank the DESY
directorate for their strong support and encouragement.
 
 \section{Appendix:                    
           Parametrisation of the feed-down contributions}
\label{sec-app}
 
Let us consider the  decay chain $D_{1,2}\rightarrow D^{*}\pi,~D^{*} \rightarrow D\pi^{0}$
in the $D^{*}$ centre-of-mass system.  Here $D_{1,2}$ is a neutral (positively charged) excited charm meson $D_1$ or $D^*_2$,
$D^{*}$ is a positively charged (neutral) $D^*$, $\pi$ is a negatively (positively) charged pion and $D$ is a  
positively charged (neutral) $D$ (charge conjugation is implied). In this system $D_{1,2}$ and $\pi$ in the initial decay and
     $D$ and $\pi^{0}$ in the subsequent decay are produced with back-to-back momenta.         
The momenta of particles in this system are:
$$
P_{\pi}^2=\left({\frac{{M^2-M_{D^{*}}^2-M_{\pi}^2}}{{2M_{D^{*}}}}}\right)^2-M_{\pi}^2,
$$    
where                                               $M$ is the $D_{1,2}$ mass;
$$
P_{D}^2=P_{\pi^{0}}^2=\left({\frac{M_{D^{*}}^2-M_{D}^2+M_{\pi^{0}}^2}{2M_{D^{*}}}}\right)^2-M_{\pi^{0}}^2
.$$ 
The measured $M(D\pi)$ is given by  
$$
M_m^2=M^2(D\pi)=M_{D}^2+M_{\pi}^2+2\sqrt{(P_{D}^2+M_{D}^2)(P_{\pi}^2+M_{\pi}^2)}-2P_{D}P_{\pi}\cos\alpha
,$$ where $\alpha$ is
    the helicity angle between $\pi^{0}$ and $\pi$.                                                 
Using the equations above, $M_m$ can be parametrised as:
\begin{equation}
M_m^2= M^2(1-a)  +b+g\sqrt{(M^2-d_1)(M^2-d_2) }\cos\alpha ,
\label{feed-downs.main}
\end{equation}
where 
$$
a=({M_{D^{*}}}^2+{M_{\pi^{0}}}^2-{M_{D}}^2)/(2{M_{D^{*}}}^2),
$$
$$
b={M_{\pi^{0}}}^2-({M_{D^{*}}}^2-{M_{\pi}}^2)({M_{D^{*}}}^2+{M_{\pi^{0}}}^2-{M_{D}}^2)/(2{M_{D^{*}}}^2),
$$
$$
g=\sqrt{({M_{D^{*}}}^2+ {M_{\pi^{0}}}^2-{M_{D}}^2)^2
-4{M_{D^{*}}}^2{M_{\pi^{0}}}^2}/(2{M_{D^{*}}}^2),
$$
$$
d_1={(M_{D^{*}}+M_{\pi})}^2,
$$
$$
d_2={(M_{D^{*}}-M_{\pi})}^2
.$$
 From Eq.(\ref{feed-downs.main}), $M$ is obtained as a function of $M_m$ and $\alpha$
$$
M=M(M_m,\alpha)
.$$ 
If the spectrum shape of $M$ is 
$$
{\frac{dN}{dM}}=f(M),$$ where $N$ is the number of candidates,
   then the $M_m$ spectrum shape   is 
$$
{\frac{dN}{dM_m}}=f(M(M_m)){\frac{dM}{dM_m}}
.$$
Combining       Eq.(\ref{feed-downs.main}) with the normalised helicity angular distribution
$$
{\frac{{dN}}{{d(\cos\alpha)}}} = {\frac{1+h\cos^2\alpha}{2(1+h/3)}},
$$
yields
$$
{\frac{{d^2 N}}{{dM_m d(\cos\alpha)}}}={f(M(M_m,\alpha)){\frac{dM}{dM_m}} {\frac{1+h\cos^2\alpha}{2(1+h/3)}}}
\label{feed-downs.full}
.$$
The fit uses the integral over $\cos\alpha$ 
\begin{equation}
{\frac{dN}{dM_m}} ={\int_{-1}^{1}f(M(M_m,\alpha)){\frac{dM}{dM_m}}{\frac{1+h\cos^2\alpha}{2(1+h/3)}}d(\cos\alpha).}
\label{feed-downs.infits}
\end{equation}
Here $f(M)$ is parametrised by                         a relativistic Breit-Wigner function as for the prompt signals.
 
For the description of the $D^{0}\pi$ spectrum, the $D^{*0} \rightarrow D^{0}\gamma$ decay was also taken into account by
replacing $M_{\pi^{0}}$ with $M_{\gamma} = 0$ in the equations above. 
For the description of the $D^{+}\pi$ spectrum, the contribution
 of the $D^{*+}\rightarrow D^{+}\gamma$  decay was neglected~\cite{PDG11}.

{\bibliographystyle{./DESY-12-144}{\raggedright\bibliography{DESY-12-144.bib}}}\vfill\eject  
\tableone
\tabletwo
\tablethree
\tablesix
\tableseven
\tablefour
\tablefive
\clearpage
\figureone
\figuretwo
\figurethree
\figurefour
\figurefive
\figuresix
\end{document}